\documentclass[11pt]{article}
 \pdfoutput = 1
 
 \usepackage[utf8]{inputenc}
 \usepackage{color,graphicx}
 \usepackage{epsfig}
 \usepackage{amsmath}
 \usepackage{verbatim}
 \usepackage{amssymb}
 \usepackage{mathabx}
 \usepackage[mathcal]{eucal}
 \usepackage{stmaryrd}
 \usepackage{empheq}
 \usepackage{esint}
 \usepackage{physics}
 \usepackage{xcolor}
 \usepackage{amsfonts}
 \usepackage{cite}
 \usepackage{array}
 \usepackage{setspace}
 \usepackage{braket}
 \usepackage{float}
 \usepackage{url}
 \usepackage{mathtools}
 \usepackage{tensor}
 \usepackage{bbold}
 \usepackage{slashed}
 \usepackage{tikz}
 \usetikzlibrary{decorations}
 \usepackage{graphicx}
 \usepackage{epstopdf}
 \usepackage{subcaption}
 \usepackage[labelfont=bf,font={sf}]{caption}
 \usepackage{pgfplots}
 \usepackage{mathrsfs}
 \usepackage{fancybox}
 \usepackage{eurosym}
 \usepackage{tcolorbox}
 \usepackage{tensor}
 \usepackage[normalem]{ulem}
 \allowdisplaybreaks
 \usepackage{changepage}

 \usepackage[margin = 2.2cm]{geometry}
 \usepackage[ragged]{footmisc}
 \usepackage[bookmarks=true,bookmarksnumbered=false,
 hyperindex=true,bookmarksopen=true,hyperfigures=true,
 colorlinks=true,linkcolor=webblue,citecolor=webgreen,urlcolor=webblue,breaklinks]{hyperref}
 \definecolor{webgreen}{rgb}{0, 0.5, 0}
 \definecolor{webblue}{rgb}{0, 0, 0.5}
 \definecolor{webred}{rgb}{0.5, 0, 0}
 \definecolor{darkgreen}{rgb}{0,0.5,0}

 \newcommand{\average}[1]{\left\langle #1 \right\rangle}

 \def\ben{\begin{equation}}
 \def\een{\end{equation}}

    \let\d=\delta 
    
      \let\r=v

 \def\be{\begin{equation}}
 \def\ee{\end{equation}}
 \def\ba{\begin{array}}
 \def\ea{\end{array}}

 \def\dalemb#1#2{{\vbox{\hrule height .#2pt
 \hbox{\vrule width.#2pt height#1pt \kern#1pt
 \vrule width.#2pt}
 \hrule height.#2pt}}}
 
 \newcommand{\bea}{\begin{eqnarray}}
 \newcommand{\eea}{\end{eqnarray}}

 \def\Im{{{\mathfrak{Im}}}}

 \let\tilde=\widetilde

 \renewcommand{\d}{\mathrm{d}}
 \renewcommand{\i}{\mathrm{i}}

 \definecolor{Blue}{HTML}{0072B2}
\definecolor{Orange}{HTML}{E69F00}
\definecolor{Green}{HTML}{66B300}

 \numberwithin{equation}{section}
 \newcommand{\obscolor}{Orange}
 \newcommand{\bracolor}{Blue}
 \newcommand{\bradiffcolor}{Green}
 
\begin{document}
 
\thispagestyle{empty}
 ~\vspace{5mm}
\begin{adjustwidth}{-1cm}{-1cm}
\begin{center}
 {\LARGE \bf 
Absolute entropy and the observer's no-boundary state}
 \vspace{0.4in}

 {\bf Andreas Blommaert${}^1$, Jonah Kudler-Flam${}^{1,2}$, and Erez Y. Urbach${}^1$}
 \end{center}
 \end{adjustwidth}
\begin{center}
 \vspace{0.4in}
 {${}^1$School of Natural Sciences, Institute for Advanced Study, Princeton, NJ 08540, USA}

 {${}^2$Princeton Center for Theoretical Science, Princeton University, Princeton, NJ 08544, USA}
 \vspace{0.1in}
 
 {\tt blommaert@ias.edu, jkudlerflam@ias.edu, urbach@ias.edu}
\end{center}
 
 \vspace{0.4in}
 
 \begin{abstract}
 \noindent We investigate the no-boundary proposal for closed universes with an observer. We argue that the observer's no-boundary state is the identity operator on the physical Hilbert space, i.e., the maximum entropy state and show this explicitly in Jackiw-Teitelboim gravity. Geometrically, the no-boundary state is a bra-ket wormhole.
 Expectation values in the no-boundary state provide a trace for the observer's algebra, which allows one to define von Neumann entropy for observers in different universes as the relative entropy with respect to the no-boundary state. This result is consistent with all previously discussed cases of traces for invariantly defined regions.
 \end{abstract}

 \pagebreak
 \setcounter{page}{1}
 \setcounter{tocdepth}{2}
 \tableofcontents

\section{Introduction}\label{sect:1.intro}

The advent of the AdS/CFT correspondence \cite{Maldacena:1997re, Gubser:1998bc, Witten:1998qj,Aharony:1999ti} has led to tremendous progress in our understanding of quantum gravity in open spacetimes. Much less is known about quantum gravity in closed spacetimes. One reason for this is that because diffeomorphisms are redundant in gravity, to describe local physics, one must consider ``relational observables'' \cite{DeWitt:1967yk}. In open universes, bulk observables can be defined with respect to an asymptotic holographic boundary. In the absence of such a boundary, such as for a closed universe or black hole interiors, it is less obvious how to define meaningful bulk observables. This has stymied progress.

A way forward is to ``dress'' observables to other features of the spacetime, such as an observer \cite{Page:1983uc}. Recently, Witten proposed a general construction of quasi-local gauge-invariant observables dressed to an observer's worldline \cite{witten2024background}, extending the construction in de Sitter space with Chandrasekaran, Longo, and Penington (CLPW) \cite{Chandrasekaran:2022cip}. 
The ``observer's algebra'' captures the experiences of the observer in perturbation theory around a given background geometry. In this paper, we combine this progress with a recent improved understanding of the gravitational path integral for closed universes. 

An important application of the gravitational path integral for closed universes is the no-boundary proposal \cite{hartle1983wave}. It proposes that the state of the universe is prepared by the gravity path integral with no boundaries, aside from the boundary where measurements are performed. In this paper, we introduce the ``observer's no-boundary state.'' Our main result is that this no-boundary density matrix is the identity operator (once properly defined) for general closed universes under modest assumptions, establishing a proposal by Witten \cite{witten2024background}
\begin{equation}
 \rho_\text{NB}=\mathbb{1}\,.\label{1.1main}
\end{equation}
Our arguments go beyond perturbation theory, and follow directly from the gravitational path integral and an interpretation of the no-boundary proposal. Before presenting intuition for the validity of \eqref{1.1main}, we recall why this proposal is useful to define entropies.

Following the great success of studying von Neumann entropies in open universes
\cite{Ryu:2006bv,Ryu:2006ef,Hubeny:2007xt,Lewkowycz:2013nqa,Faulkner:2013ana, Engelhardt:2014gca,Penington:2019npb,Almheiri:2019psf,Almheiri:2019hni,Penington:2019kki,Almheiri:2019qdq}, one would like to assign entropies to states in the observer's Hilbert space in closed universes. For the static patch of de Sitter space, CLPW successfully calculated entropy differences between two perturbative states using the observer's algebra, and found agreement with the generalized entropy of the static patch. This construction was later generalized to several other invariantly defined regions with horizons \cite{Chandrasekaran:2022eqq, Kudler-Flam:2023qfl, Kudler-Flam:2024psh, Chen:2024rpx} as well as proposals for treating more general subregions in semiclassical quantum gravity \cite{Jensen:2023yxy, Chen:2024rpx}. The observable algebra in these constructions is classified as a type II von Neumann algebra. Type II algebras allow for a calculation of entropy differences unambiguously.\footnote{See recent reviews on von Neumann algebra application in quantum-information theory in quantum field theory \cite{Witten:2021jzq} or their type classification \cite{Sorce:2023fdx}.} By their perturbative nature, unfortunately, these constructions leave undetermined a state-independent, but background dependent, leading $O(1/G)$ contribution to entropy
so do not enable one to compute entropy differences between two different backgrounds or for the calculation of the absolute entropy.

To address this issue, Witten suggested an intriguing hypothesis \cite{witten2024background} 
motivated by the standard finite-dimensional identity 
\begin{equation}\label{eq:S_S_rel}
 S_\text{vN}(\rho) = -S_\text{rel}(\rho \parallel \mathbb{1})\,,
\end{equation}
between the von-Neumann entropy and the relative entropy with the identity matrix.
Given an understanding of the (unnormalized) identity matrix, instead of defining von Neumann entropy directly, one can define it as relative entropy with respect to the identity operator. This is convenient because relative entropies are generally well-defined, even in infinite-dimensional systems such as quantum field theory.\footnote{Relative entropy between two states can be defined for any von Neumann algebra \cite{araki1975relative,araki1977relative}. More generally, it can be defined for $C^*$ algebras using an appropriate algebraic definition of the states \cite{Uhlmann:1976me,Belavkin:1982}.} This reduces the question of defining the absolute entropy $S_\text{vN}(\rho)$ to an understanding of the (unnormalized) quantum gravity state that is given by $\mathbb{1}$, which is proposed to be
the no-boundary state, \eqref{1.1main}. In the context of de Sitter spacetimes, this suggestion is consistent with the long-standing lore that the no-boundary state 
for the
static patch has maximal entropy \cite{Maeda:1997fh,Bousso:2000nf,Bousso:2000md,Dong:2018cuv}. Our main result is establishing this proposal more generally, under standard assumptions about the gravitational path integral.

In summary, the identification \eqref{1.1main} allows one to define the absolute entropy of states including in particular the $O(1/G)$ contribution via an asymptotic expansion, whereas previously only $O(1)$ entropy differences were available using algebraic methods. We briefly summarize the remainder of the paper.

In \textbf{section \ref{sect:2.obscomplete}}, we consider spacetimes in which the observer has access to a complete Cauchy slice. The matrix elements of $\rho_\text{NB}$ are then between two global slices of the universe.
We will argue that the leading topology that contributes to $\rho_\text{NB}$ is a wormhole connecting the bra and ket
\begin{equation}\label{eq:NB_global}
 \rho_\text{NB}=\ \mathrel{\raisebox{-0.5\height}
 {\begin{tikzpicture}
        \draw[thick, \bracolor] (0,0) ellipse (1 and .25);
    \draw[thick, \bracolor] (0,-1) ellipse (1 and .25);
    \draw[thick ] (-1,0) to[out = 90, in = 90] (-1.5,-.5) to[out = -90, in = -90](-1,-1) ;
    \draw[thick ] (1,0) to[out = 90, in = 90] (-2,-.5) to[out = -90, in = -90](1,-1) ;
    \draw[thick ,\obscolor] (0,.25) to[out = 90, in = 90] (-1.75,-.5) to[out = -90, in = -90](0,-1.25) ;
    \fill[\obscolor] (0,+.25) circle (2pt);
    \fill[\obscolor] (0,-1.25) circle (2pt);
\end{tikzpicture}}}
\end{equation}
where the observer's worldline is drawn in orange and the Cauchy slices in blue.
Crucially, the leading no-boundary topology (two hemispheres) is forbidden by the connectivity of the observer worldline. This is in line with recent discussions of the rules for the gravitational path integral in the presence of an observer \cite{lucaobserver,Akers:2025ahe, Harlow:2025pvj}. We will present a general path integral argument for $\rho_\text{NB}=\mathbb{1}$. Intuitively, this path integral gives the identity operator because time evolution in gravity acts trivially, as the Hamiltonian is a constraint $H_{\text{WDW}}=0$.

In \textbf{section \ref{sect:4.JT}}, we check the somewhat formal arguments of section \ref{sect:2.obscomplete} by showing that, for closed universes in Jackiw-Teitelboim (JT) gravity with $\Lambda < 0$, the path integral that prepares the observer's no-boundary state \eqref{eq:NB_global} indeed matches with matrix elements of the identity operator $\mathbb{1}$ on the physical Hilbert space.

We believe that \eqref{eq:NB_global} is not only consistent with, but is actually the natural extension of the sphere preparation of the no-boundary state for the de Sitter static patch with an observer \cite{Chandrasekaran:2022cip}, schematically drawn by
\begin{equation}\label{eq:NB_region}
 \rho_\text{NB}= \ \mathrel{\raisebox{-0.5\height}{\begin{tikzpicture}
    \draw[thick, \bracolor] (0,0 ) -- (1,1)--(2,0) --(1,-1)--cycle;
     \fill[\obscolor] (1,1) circle (2pt); 
     \fill[\obscolor] (1,-1) circle (2pt); 
     \fill[] (0,0) circle (2pt);
      \fill[] (2,0) circle (2pt);
    \draw[thick] (1,0) circle (1.5);
    \draw[thick, \obscolor] (1,1) to[out = 90,in= 90](-.25,0) to[out=-90,in=-90] (1,-1); 
\end{tikzpicture}}}
\end{equation}
The observer's observations are limited by a causal horizon, so the matrix elements of $\rho_\text{NB}$ are between two Cauchy slices of the causal diamond (blue). We use the null boundaries of the diamond in order to have an invariantly defined subregion related to the observer in more general spacetimes, as will become clear in \textbf{section \ref{sect:3obshorizon}}. Semiclassically, this path integral prepares the Bunch-Davies state for the static patch. In both \eqref{eq:NB_global} and \eqref{eq:NB_region}, the path integrals describe the leading topology for the observer's no-boundary proposal. However, as in the de Sitter case, the observer causal access is bounded by a horizon, the leading topology is a sphere. One can interpret \eqref{eq:NB_region} as the ``partial trace'' of \eqref{eq:NB_global} to a properly defined region.
{Section \ref{sect:3obshorizon}} is devoted to generalizing the path integral arguments of section \ref{sect:2.obscomplete} to invariantly defined subregions of closed spacetimes, which naturally reproduces and extends results from \cite{Chandrasekaran:2022cip} and \cite{Chen:2024rpx} to more general cases where the observer has causal horizons.

In principle, it should be possible to explicitly check our claim for the case of an observer with causal horizons in the perturbative $G\to 0$ limit. One would like to consider examples of such spaces that are not pure de Sitter (dS), construct the observer's algebra and check that the tracial state is indeed prepared by the no-boundary path integral. So, one would like to generalize the construction of \cite{Chandrasekaran:2022cip, witten2024background} beyond dS. One key ingredient in their construction was the KMS property of correlation functions on the worldline of the observer. In \textbf{section \ref{sect:nogo}}, we show there is no KMS state for worldline correlation functions for any asymptotically de Sitter geometry besides de Sitter itself. We speculate that the notion of ``weak KMS'' which is generally satisfied by worldline correlation functions 
prepared by a no-boundary geometry 
might, however, suffice to define the trace explicitly.

In the discussion (\textbf{section \ref{sect:5.conclusion}}) we discuss intriguing classical FLRW wormhole solutions that contribute to the no-boundary density matrix. We believe that these are a good testing ground for the ideas of sections \ref{sect:3obshorizon} and \ref{sect:nogo}. We point out new questions that these solutions pose.

\section{Observers with complete causal access}\label{sect:2.obscomplete}
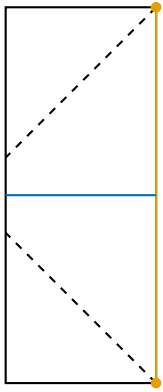
\begin{figure}
 \centering
 \begin{tikzpicture}
    \draw[thick] (0,0) -- (2,0) -- (2,5)--(0,5)--cycle ;
    \draw[thick,dashed] (2,0) -- (0,2);
    \draw[thick,dashed] (2,5) -- (0,3);
    \draw[thick, \obscolor] (2,5) --(2,0);
    \draw[thick,\bracolor] (0,2.5)  -- (2,2.5);
    \fill[\obscolor] (2,0) circle (2pt); 
     \fill[\obscolor] (2,5) circle (2pt); 
\end{tikzpicture}
 \caption{Penrose diagram for an observer (orange) with complete access to a Cauchy surface (blue). Explicit solutions with this Penrose diagram are described in section \ref{sect:5.conclusion}.}
 \label{fig:complete1}
\end{figure}

In this section, we consider the observer's no-boundary state in the case where the observer has complete causal access. For example, consider a semiclassical spacetime where an observer, whose worldline extends infinitely to the future and past, has causal access to a complete Cauchy slice in a closed universe such that the Penrose diagram is as in figure \ref{fig:complete1}. Because the observer sees the global quantum state at some time, they have complete information about the quantum state at all times by Schr\"odinger evolution. The observer's algebra will be type I because their exist pure states.

The main goal of this section is to argue that the gravitational path integral prepares the identity matrix on a complete Cauchy slice when imposing the no-boundary rules, \eqref{1.1main}. We first detail why this occurs while ignoring the effects of topology change. In section \ref{sect:2.1}, we explain that the observer's no-boundary state is a mixed state and gets its leading contribution from geometries directly connecting the bra and ket boundary conditions. Crucially, the usual Hartle-Hawking no-boundary geometry does not contribute. In section \ref{sect:2.2}, we then show explicitly that $\rho_\text{NB}=\mathbb{1}$. The basic point is that the gravitational path integral that computes matrix elements of $\rho_\text{NB}$ is the same path integral that computes inner products. In section \ref{sec:canonical_quant}, we show that the Arnowitt–Deser–Misner (ADM) formalism leads to the same conclusion. In section \ref{sect:2.4topo}, we explain that this property is not affected by taking into account topology change. The arguments in this section are general, and so necessarily somewhat abstract. We complement this by working out the details explicitly for JT gravity in section \ref{sect:4.JT}.

\subsection{Nothing but bra-ket wormholes}\label{sect:2.1}
Matrix elements of the no-boundary state are specified by the induced metric, the configuration of the quantum fields, and the observer on a Cauchy slice. We will collectively denote these boundary conditions as $\phi$.\footnote{In a theory without time-reversal symmetry, one may also want to impose an orientation on the manifold to specify whether the boundary represents a bra or ket \cite{Witten:2025ayw}.} One then sums (path integrates) over all configurations with no additional spacetime boundaries. 
The action of the observer taken to be
\begin{align}
 I_\text{obs} =\int_{\gamma}\d\tau\left(p\frac{\d q}{\d\tau}-\sqrt{-g_{\tau \tau}}(m+q)\right),
\end{align}
where $\gamma$ is the worldline parametrized by $\tau$ and the mass $m$ is large in cosmological units buts small in Planck units so that the geodesic approximation is valid without significant backreaction. $q>0$ is the observer's energy, which could have (in the idealized limit) any non-negative real value.

Our first observation is that for the observer's no-boundary state, where one considers Cauchy slices with one observer present, there are no ``no-boundary'' field configurations of the usual type imagined by Hartle and Hawking, where spacetime caps off in the (complex) past of the Cauchy slice. Indeed, without a second boundary, the observer worldline has no place to end
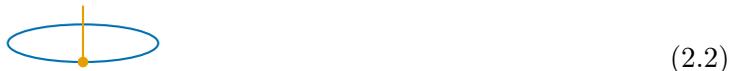
\begin{align}
    \begin{tikzpicture}
    \draw[thick,\bracolor] (0,0) ellipse (1 and .25);
    \fill[\obscolor] (0,-.25) circle (2pt);
    \draw[thick,\obscolor] (0,-.25) -- (0,.5);
    \end{tikzpicture}
\end{align}
In order to make sense of the observer's no-boundary state, we propose that one must instead consider the density matrix, rather than only consider pure states. In a density matrix, one specifies boundary field values for both the bra and the ``ket,'' as follows 
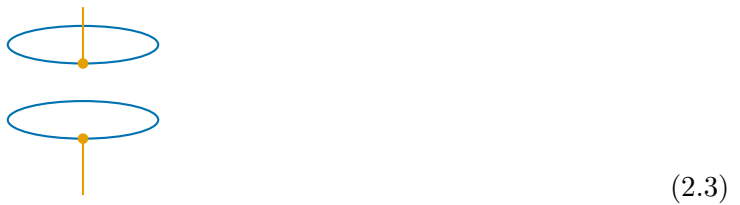
\begin{align}
    \begin{tikzpicture}
    \draw[thick,\bracolor] (0,0) ellipse (1 and .25);
    \draw[thick,\bracolor] (0,-1) ellipse (1 and .25);
    \fill[\obscolor] (0,-.25) circle (2pt); 
    \fill[\obscolor] (0,-1.25) circle (2pt); 
    \draw[thick,\obscolor] (0,-.25) -- (0,.5); 
    \draw[thick,\obscolor] (0,-1.25) -- (0,-2); 
\end{tikzpicture}
\end{align}
There now exist configurations where the observer worldline stretches from the bra to the ``ket.'' The connection between the bra and ket implies that the no-boundary state is a mixed state. Such bra-ket wormhole configurations were previously considered as contributions to the no-boundary density matrix \cite{page1986density,hawking1987density,Chen:2020tes}. The important distinction with our setup is that, for the observer's density matrix, the usual Hartle-Hawking disconnected contribution never exists. The connected (wormhole) solution always provides the leading contribution to the path integral. 

In summary, the observer's no-boundary state is computed by the path integral over spacetimes with wormhole topology $\mathcal{M}\times I$ and the specified boundary conditions. Denoting the bulk metric, quantum fields, and observer data by $\phi_\text{bulk}$, matrix elements of $\rho_\text{NB}$ are schematically computed as
\begin{align}
\label{eq:nb_density_filled}
\bra{\phi''}\rho_{NB}\ket{\phi'} = \int \mathcal{D}\phi_\text{bulk}\quad \raisebox{-0.47\height}{\begin{tikzpicture}
    \draw[thick, \bracolor] (0,0) ellipse (1 and .25);
    \draw[thick, \bracolor] (0,-1) ellipse (1 and .25);
    \fill[\obscolor] (0,.25) circle (2pt);
    \fill[\obscolor] (0,-1.25) circle (2pt);
    \draw[thick ] (-1,0) to[out = 90, in = 90] (-1.5,-.5) to[out = -90, in = -90](-1,-1) ;
    \draw[thick ] (1,0) to[out = 90, in = 90] (-2,-.5) to[out = -90, in = -90](1,-1) ;
    \draw[thick ,\obscolor] (0,.25) to[out = 90, in = 90] (-1.75,-.5) to[out = -90, in = -90](0,-1.25) ;
    \node at (-2.7,-.5) {$\phi_\text{bulk}$};
    \node at (1.5,0) {$\phi''$};
    \node at (1.5,-1) {$\phi'$};
\end{tikzpicture}}.
\end{align}
Contributions from spacetimes with topology change will be included in section \ref{sect:2.4topo}. We have not specifed whether the path integral is over Lorentzian, Euclidean, or complex spacetimes connecting the bra and ket because our conclusions to not rely on this, only on the assumption that there does exist a well-defined contour of integration.
Several contours have been considered for bra-ket wormholes in two-dimensional gravity in \cite{Chen:2020tes,Fumagalli:2024msi}.
A simple way of thinking about bra-ket wormholes is to consider a purely Lorentzian geometry that connects two time slices of the same classical geometry. The data on one of these slices represents the bra boundary condition and the data on the other time slice represents the ket boundary conditions. This Lorentzian way of thinking about bra-ket wormholes is directly related with our Hamiltonian discussion in section \ref{sec:canonical_quant}. In section \ref{sect:4.JT}, we demonstrate the equivalence of the Lorentzian and Euclidean descriptions of bra-ket wormholes in JT gravity.

\subsection{The no-boundary state is the identity}
\label{sect:2.2}
With the path integral definition of the observer's no-boundary state \eqref{eq:nb_density_filled}, we would now like to argue that $\rho_\text{NB}=\mathbb{1}$. The simplest way to argue this is to think about the inner product in quantum gravity. A natural candidate is the inner product induced by the gravitational path integral \cite{hartle1983wave}, filling in the bulk region in all possible ways that has the prescribed states as boundary conditions
\begin{align}
 \bra{\phi''}{\phi'}\rangle = \int \mathcal{D}\phi_\text{bulk}\,\,\,\,\raisebox{-0.46\height}{\begin{tikzpicture}
    \node at (-1.7,-1) {$\phi_\text{bulk}$};
    \node at (1.5,0) {$\phi''$};
    \node at (1.5,-2) {$\phi'$};
    \draw[thick,\bracolor] (0,0) ellipse (1 and .25);
    \draw[thick,\bracolor] (0,-2) ellipse (1 and .25);
    \fill[\obscolor] (0,-.25) circle (2pt);
    \fill[\obscolor] (0,-2.25) circle (2pt);
    \draw[thick,\obscolor] (0,-.25) -- (0,-1.5);
    \draw[thick,\obscolor] (0,-2.25) -- (0,-1.5);
    \draw[thick] (-1,0)to[out = -70,in = 80]  (-1,-1.5) to[out = -100,in = 110]  (-1,-2) ;
     \draw[thick] (1,0)to[out = -70,in = 80]  (1,-1.5) to[out = -100,in = 110]  (1,-2) ;
\end{tikzpicture}}.
\end{align}
In particular, one generally only includes geometries in the inner product that have no other boundaries.
Since in $\phi_\text{bulk}$ one path integrates over all spacetimes with the relevant boundary conditions, we see that this is the same equation as for the matrix elements $\bra{\phi''}\rho_\text{NB}\ket{\phi'}$. Therefore 
\begin{equation}
 \rho_\text{NB}=\mathbb{1}.
\end{equation}
Using the gravitational path integral as an inner product is fairly uncontroversial. Therefore, we believe the same is true for the statement that the observer's no-boundary state is the identity matrix.

It is important to explain on which Hilbert space this is the identity matrix. In quantum gravity, we must distinguish the space of boundary conditions from the physical Hilbert space \cite{Witten:2022xxp}. The physical Hilbert space is labeled (roughly speaking) by the space of boundary conditions modulo diffeomorphisms. This includes time-like diffeomorphisms generated by the Wheeler-DeWitt (WDW) Hamiltonian constraint \cite{DeWitt:1967yk}, which evolves between different Cauchy slices in the same geometry. The observer's no-boundary state is the identity on the physical Hilbert space which we label $\ket{\psi}$. As we have explained, there is an equivalence class of boundary conditions $\psi$ (related by diffeomorphisms) associated with $\ket{\psi}$. These statements will be further clarified in section \ref{sec:canonical_quant}.

To drive home the point that $\rho_\text{NB}=\mathbb{1}$, we will investigate the square of the density matrix. In quantum field theory path integrals, matrix multiplication is achieved by gluing the ket of one matrix to the bra of the other by path integrating over boundary conditions $\phi$. The situation is more subtle in quantum gravity. Instead of gluing by integrating over all boundary conditions, one should only integrate over boundary conditions modulo diffeomorphisms. This reflects matrix multiplication in the physical Hilbert space by avoiding overcounting which could arise, for instance, by gluing the same geometry on different Cauchy slices (which are related by gauge transformations) \cite{Witten:2022xxp}. Matrix elements of the square of the observer's no-boundary state are therefore computed as\footnote{The path integral over boundary conditions modulo diffeomorphisms $\psi$ comes with a nontrivial Faddeev–Popov determinant in any particular gauge choice which we leave implicit \cite{Witten:2022xxp}. We will be more precise in section \ref{sect:3.1HJT}.}
\begin{equation}
 \bra{\phi''}\rho_\text{NB}^2\ket{\phi'}=\underset{\substack{\text{boundary conditions}\\\text{modulo diffeomorphisms}}}{\int\mathcal{D}\psi}\, \bra{\phi''}\rho_\text{NB}\ket{\psi}\bra{\psi}\rho_\text{NB}\ket{\phi'}\,.
\end{equation}
Omitting the integration over bulk fields $\phi_\text{bulk}$ (to avoid clutter), this can be graphically represented as
\begin{align}
\label{eq:nb_density_squared}
 \bra{\phi''}\rho_\text{NB}^2\ket{\phi'} = \underset{\substack{\text{boundary conditions}\\\text{modulo diffeomorphisms}}}{\int\mathcal{D}\psi}\quad \raisebox{-0.46\height}{\begin{tikzpicture}
    \draw[thick, \bracolor] (0,0) ellipse (1 and .25);
    \draw[thick, \bradiffcolor] (0,-1) ellipse (1 and .25);

    \draw[thick ] (-1,0) to[out = 90, in = 90] (-1.5,-.5) to[out = -90, in = -90](-1,-1) ;
    \draw[thick ] (1,0) to[out = 90, in = 90] (-2,-.5) to[out = -90, in = -90](1,-1) ;
    \draw[thick ,\obscolor] (0,.25) to[out = 90, in = 90] (-1.75,-.5) to[out = -90, in = -90](0,-1.25) ;
    \node at (1.5,0) {$\phi''$};
    \node at (1.5,-1) {$\psi$};
    \draw[thick,\bradiffcolor] (0+4.5,0) ellipse (1 and .25);
    \draw[thick, \bracolor] (0+4.5,-1) ellipse (1 and .25);
    \draw[thick ] (-1+4.5,0) to[out = 90, in = 90] (-1.5+4.5,-.5) to[out = -90, in = -90](-1+4.5,-1) ;
    \draw[thick ] (1+4.5,0) to[out = 90, in = 90] (-2+4.5,-.5) to[out = -90, in = -90](1+4.5,-1) ;
    \draw[thick ,\obscolor] (0+4.5,.25) to[out = 90, in = 90] (-1.75+4.5,-.5) to[out = -90, in = -90](0+4.5,-1.25) ;
    \node at (1.5+4.5,0) {$\psi$};
    \node at (1.5+4.5,-1) {$\phi'$};

        \fill[\obscolor] (0,.25) circle (2pt);
    \fill[\obscolor] (0,-1.25) circle (2pt);
    \fill[\obscolor] (4.5,.25) circle (2pt);
    \fill[\obscolor] (4.5,-1.25) circle (2pt);
\end{tikzpicture}},
\end{align}
where the green slices represent boundary conditions modulo diffeomorphisms in contrast to the blue slices which are any fixed boundary conditions. 
Hence, the prescription in the gravitational path integral to compute $\bra{\phi''}\rho_\text{NB}^2\ket{\phi'}$ is to path integrate over all configurations with wormhole topology $\mathcal{M}\times I$ with boundary conditions $\phi'$ and $\phi''$. This is identical to the computation of $\bra{\phi''}\rho_\text{NB}\ket{\phi'}$ in \eqref{eq:nb_density_filled}. Therefore
\begin{equation}
 \rho_\text{NB}^2=\rho_\text{NB}.
\end{equation}
This leads to the conclusion that the no-boundary density matrix is a projector on the physical Hilbert space, i.e.~$\rho_{\text{NB}} =\mathbb{1}$. 

It is instructional to compare the squaring of $\rho_{\text{NB}}$ to the squaring of similar-looking path integrals in non-gravitational quantum field theories. As a simple example, consider the path integral in a quantum field theory on a manifold $\mathcal{M}\times [0,\beta]$. This density matrix is a Gibbs state at inverse temperature $\beta$
\begin{align}
 \bra{\phi''}e^{-\beta H}\ket{\phi'} = \int\mathcal{D}\phi_\text{bulk}\quad \mathrel{\raisebox{-0.5\height}{\begin{tikzpicture}
    \node at (-1.6,-0.5) {$\phi_\text{bulk}$};
    \node at (-1.5,0) {$\phi''$};
    \node at (-1.5,-1) {$\phi'$};
    \draw[thick,\bracolor] (0,0) ellipse (1 and .25);
    \draw[thick,\bracolor] (0,-1) ellipse (1 and .25);
    \draw[thick] (-1,0)  --(-1,-1) ;
     \draw[thick] (1,0)--  (1,-1) ;
     \node at (1.5,-0.5) {$\beta$};
\end{tikzpicture}}}.
\end{align}
The square of this density matrix is the Gibbs state at inverse temperature $2\beta$, $e^{-2\beta H}$
\begin{align}
 \bra{\phi''}e^{-2\beta H}\ket{\phi'} = \int\mathcal{D}\phi_\text{bulk}\quad \mathrel{\raisebox{-0.5\height}{\begin{tikzpicture}
    \node at (-1.6,-1) {$\phi_\text{bulk}$};
    \node at (-1.5,0) {$\phi''$};
    \node at (-1.5,-2) {$\phi'$};
    \draw[thick,\bracolor] (0,0) ellipse (1 and .25);
    \draw[thick,\bracolor] (0,-2) ellipse (1 and .25);
    \draw[thick] (-1,0)  --(-1,-2) ;
     \draw[thick] (1,0)--  (1,-2) ;
     \node at (1.5,-1) {$2\beta$};
\end{tikzpicture}}}.
\end{align}
This is clearly not a projector unless $H=0$. One case where $H = 0$ (besides gravity) is topological quantum field theory. 
The analogy between gravity and standard topological quantum field theory however no longer holds when including topology change, see section \ref{sect:2.4topo}. 

The norm of the no-boundary state is formally given by the trace
\begin{align}
 \Tr \rho_\text{NB} = \underset{\substack{\text{boundary conditions}\\\text{modulo diffeomorphisms}}}{\int\mathcal{D}\psi}\int \mathcal{D}\phi_\text{bulk}\,\,\,\raisebox{-0.47\height}{\begin{tikzpicture}
    \draw[thick, \bradiffcolor] (0,0) ellipse (1 and .25);
    \draw[thick,\bradiffcolor] (0,-1) ellipse (1 and .25);
    \fill[\obscolor] (0,.25) circle (2pt);
    \fill[\obscolor] (0,-1.25) circle (2pt);
    \draw[thick ] (-1,0) to[out = 90, in = 90] (-1.5,-.5) to[out = -90, in = -90](-1,-1) ;
    \draw[thick ] (1,0) to[out = 90, in = 90] (-2,-.5) to[out = -90, in = -90](1,-1) ;
    \draw[thick ,\obscolor] (0,.25) to[out = 90, in = 90] (-1.75,-.5) to[out = -90, in = -90](0,-1.25) ;
    \node at (-2.7,-.5) {$\phi_\text{bulk}$};
    \node at (1.5,0) {$\psi$};
    \node at (1.5,-1) {$\psi$};
\end{tikzpicture}}.
\end{align}
This computes the gravitational path integral on $\mathcal{M}\times S_1$. In many cases, including JT gravity discussed in section \ref{sect:4.JT}, the answer is divergent. In these cases, the no-boundary state is more properly referred to a ``weight,'' which is an unnormalizable state, like a position eigenstate of the harmonic oscillator.

In cases where there is a dominant saddle point, the norm is approximately $e^{-I}$, with $I$ the on-shell action. We may evaluate the von Neumann entropy, as defined via the relative entropy, of the unit normalized observer's no-boundary state $\tilde \rho_\text{NB} \equiv \left(\text{Tr} \rho_\text{NB}\right)^{-1} \rho_\text{NB}$
\begin{align}
 S_\text{vN}(\tilde \rho_{\text{NB}}) \equiv -S_\text{rel}(\tilde \rho_{\text{NB}}\parallel\rho_\text{NB} )= -S_\text{rel}(\tilde \rho_{\text{NB}}\parallel\tilde \rho_{\text{NB}} ) -I= -I\,,
\end{align}
as well as the entropy of any normalized state $\rho$ 
\begin{align}\label{eq:rel_vn_I}
S_\text{vN}(\rho) = - S_\text{rel}(\rho  \parallel  \tilde{\rho}_{\text{NB}}) -I.
\end{align}
In section \ref{sect:5.conclusion}, we discuss an explicit example of this, an Einstein static universe. In that case, the classical action, $I$, is zero. This is consistent with there being no additive constant in states of a type I algebra.

\subsection{Lorentzian quantum gravity and the lapse contour}\label{sec:canonical_quant}
In this section we present an alternative (though more heuristic) derivation of \eqref{1.1main}, still ignoring topology change.
 In particular, we detail how to compute the inner products between boundary states $\ket{\phi}$ in quantum gravity and show that this inner product is implemented by a Lorentzian path integral over spacetimes of topology $\mathcal{M}\times I$ which we argued computes the no-boundary state \eqref{eq:nb_density_filled}.

Consider first the Lorentzian path integral in the ADM formalism \cite{Arnowitt:1959ah,Wiltshire:1995vk} (see also \cite{Barvinsky:2013nca} for relevant discussion), where the metric is parameterized in terms of the induced metric $g_{ij}$, the lapse $N$ and shift vector $N^j$
\begin{align}
 \d s^2 = -N^2 \d t^2 + g_{ij}(\d x^i + N^i \d t)(\d x^j +N^j \d t)\,,\label{2.18dsquared}
\end{align}
all functions of $t$, $x^i$.
Introducing conjugate momenta $\Pi_{i j} = -\i \delta/\delta g_{ij}$ for the induced metric one finds
the Lorentzian action simply takes the form
\begin{equation}
 I= \int \d t \d^3 x \bigg( \Pi^{ij}\dot g_{ij} -N H_\text{WDW}-\sum_i N^i P_i\bigg)\,.
\end{equation}
Here $H_\text{WDW}$ is a function on phase space called the Hamiltonian constraint
\begin{equation}
\begin{split}
 H_\text{WDW} 
 &= -R(g)+2\Lambda + \frac{(16\pi G)^2}{\det g} \left(\Pi_{ij}\Pi^{ij}-\Pi_i^i \Pi_j^j \right)+\text{matter contributions}\,, 
\end{split}
\end{equation}
with $R(g)$ the three-dimensional Ricci scalar.
Crucially, the action does not depend on time derivatives of $N$ and $N^j$ and they do not appear in the phase space. Written in a Hamiltonian form, the Lorentzian path integral on spacetimes over topology $\mathcal{M}\times I$ with boundary conditions $\phi'$ and $\phi''$ therefore becomes
\begin{equation}
 \bra{\phi''}\rho_\text{NB}\ket{\phi'}= \bra{\phi''}\frac{1}{2\pi}\int_{-\infty}^{+\infty}\d N\,e^{-\i H_\text{WDW} N }\ket{\phi'}\,.\label{2.21rigg}
\end{equation}
Here, we made the gauge choice that $t$ in \eqref{2.18dsquared} runs between $-1/2$ on one Cauchy slice and $1/2$ on the other Cauchy slice, 
and gauged-fix the lapse function to be $t$-independent (see \cite{Halliwell:1988wc} for more details). Furthermore, we consider a minisuperspace truncation where the entire metric is homogeneous (otherwise one path integrates over $N(x^j)$) and we left implicit the integrals over the shifts implementing the so-called momentum constraints $P_i=0$ in the path integral. 

Equation \eqref{2.21rigg} demonstrates that the Lorentzian no-boundary path integral implements a ``rigging map'' $\eta$ \cite{Marolf:1996gb}
\begin{equation}
 \rho_\text{NB}=\eta=\frac{1}{2\pi}\int_{-\infty}^{+\infty}\d N\,e^{-\i H_\text{WDW} N }=\delta(H_\text{WDW})\,.\label{2.22rhoisdelta}
\end{equation}
There has been discussion in the literature regarding the contour of this lapse integral, for instance \cite{DiazDorronsoro:2017hti,Feldbrugge:2017kzv,Banihashemi:2024aal}. However, the most reasonable answers are obtained when integrating $N$ from $-\infty$ to $+\infty$. One a priori reason for this contour is that this results in a Hermitian density matrix. Another reason is that, in the spirit of gauging spacetime inversions \cite{Harlow:2023hjb}, the bra and ket should be on equal footing in terms of their time ordering. In other words, we should consider Lorentzian geometries where one first encounters the data $\phi'$ at $t=-1/2$ and later the data $\phi''$ at $t=+1/2$, but also Lorentzian geometries where one encounters the data $\phi''$ at $t=-1/2$ and only later the data $\phi'$ at $t=+1/2$. This effectively means one should integrate over $N>0$ and $N<0$ in Lorentzian quantum gravity.

This may naively seem in tension with the fact that we can alternatively compute the no-boundary state using the Euclidean path integral, where one imagines integrating in some sense only over positive lapses $N>0$. To accommodate the potentially skeptical reader, we will show explicitly in section \ref{sect:4.JT} that the Lorentzian and Euclidean path integrals agree for AdS JT gravity both with and without an observer.

We now explain in which sense \eqref{2.22rhoisdelta} is consistent with our earlier claim that $\rho_\text{NB}=\mathbb{1}$. To do so, we must distinguish two Hilbert spaces relevant in quantum gravity.\footnote{In recent literature we recommend \cite{Held:2024rmg,Witten:2022xxp} for pedagogical and in depth explanations.} 
We will take the minisuperspace approximation of a single constraint $H_\text{WDW}$.
The ``auxiliary'' Hilbert space, $\mathcal{H}_0$, is roughly speaking the square-integrable functions on the set of boundary conditions, with states that we labelled $\ket{\phi}$. 
The physical Hilbert space, $\mathcal{H}$, is the set of physical states, which we label $|\psi\rangle\!\rangle$. The physical Hilbert space can be understood as obtained from $\mathcal{H}_0$ by identifying states that differ by a ``gauge transformation'' \cite{higuchi1991quantum1, higuchi1991quantum,Marolf:2008hg,Ashtekar:1995zh,Giulini:1998rk,Marolf:2000iq,Giulini:1999kc}
\begin{align}
 \ket{\psi} \sim \ket{\psi}+ H_\text{WDW} \ket{\phi}.
\end{align}
This leads to a Hilbert space of equivalence classes called co-invariants $|\psi\rangle\!\rangle$. The inner product on these co-invariants is 
given by the rigging map
\cite{higuchi1991quantum1, higuchi1991quantum,Marolf:2008hg,Ashtekar:1995zh,Giulini:1998rk,Marolf:2000iq,Giulini:1999kc}\footnote{See \cite{DeVuyst:2024uvd,DeVuyst:2024pop} for a detailed discussion on this approach and the work of CLPW.}
\begin{align}
\langle\!\langle{\psi'}|\psi\rangle\!\rangle =\bra{\psi'}\eta \ket{\psi}\,.\label{2.24ip}
\end{align}
Thus, we find that in the Lorentzian path integral, the no-boundary state \eqref{2.22rhoisdelta} implements an inner product and, from the right-hand-side of \eqref{2.24ip}, that it implements the identity operator on $\mathcal{H}$. This can be understood directly by realizing that in \eqref{2.22rhoisdelta}, $\delta(H_\text{WDW})$ is a projector onto the physical Hilbert space $\mathcal{H}$. 

Consider a basis for $\mathcal{H}_0$, $\{ \ket{E,\alpha} \}$, where $E$ is the eigenvalue of $H_\text{WDW}$ and $\alpha$ labels the rest of the quantum numbers.
The physical Hilbert space, $\mathcal{H}$, is understood as the eigenspace spanned by
$\{\ket{0,\alpha}\}$. If $a$ is a physical operator, i.e~one that commutes with $\mathcal{H}_\text{WDW}$, then its expectation value in the no-boundary state
is
\begin{equation}
\begin{split}
 \text{Tr}_{\mathcal{H}_0}\left(\rho_\text{NB} a\right) & = \sum_\alpha \int dE \bra{E,\alpha} \delta(H_\text{WDW}) a \ket{E,\alpha}\\
 &= \sum_\alpha \bra{0,\alpha} a \ket{0,\alpha} = \text{Tr}_\mathcal{H}(a).
\end{split}
\end{equation}
In the second line, we used that $a$ commutes with $H_\text{WDW}$ and thus acts diagonally on $E$ eigenstates. Comparing the first and last equality, we see that $\rho_\text{NB}$, as a density matrix in $\mathcal{H}_0$, is identified with the identity matrix 
in $\mathcal{H}$. 

\subsection{Including topology change}\label{sect:2.4topo}
Thus far we have not considered the impact of topology change. Unfortunately, we do not know how to include topology changing processes in the Hamiltonian description of section \ref{sec:canonical_quant}. However, they are naturally included in the gravitational path integral, and appear to be necessary for recovering the expected physics of quantum gravity \cite{Saad:2018bqo,sss,Penington:2019kki, Almheiri:2019qdq}. The basic argument of section \ref{sect:2.2} remains valid with topology change. 

We first note a contributions to the path integral of the forms\footnote{Even though these contributions have the same topology as \eqref{eq:nb_density_squared}, we include them in this section for historical reasons.}
\begin{align}
\raisebox{0.25\height}{\begin{tikzpicture}

    \draw[thick ,\obscolor] (0,-1.25)  to[out = 50, in = -130](4.5,.25)  ;
    \draw[thick ] (-1,-1.5) to[out = 70, in = -110](3.5,.5) ;
    \draw[thick ]  (1,-1.5) to[out = 70, in = -110](5.5,.5) ;
    
    \filldraw[white](-1,.5) to[out = -70, in = 110](3.5,-1.5)-- (5.5,-1.5)to[out = 110, in = -70](1,.5) ;
        \draw[thick, \bracolor] (0,.5) ellipse (1 and .25);
    \draw[thick, \bradiffcolor] (0,-1.5) ellipse (1 and .25);
    \draw[thick ] (-1,.5) to[out = -70, in = 110](3.5,-1.5) ;
    \draw[thick ] (1,.5) to[out = -70, in = 110](5.5,-1.5) ;

    \draw[thick ,\obscolor] (0,.25)  to[out = -50, in = 130](4.5,-1.25) ;
    
    \draw[thick,\bradiffcolor] (0+4.5,.5) ellipse (1 and .25);
    \draw[thick, \bracolor] (0+4.5,-1.5) ellipse (1 and .25);


    \fill[\obscolor] (0,.25) circle (2pt);
    \fill[\obscolor] (0,-1.25) circle (2pt);
    \fill[\obscolor] (4.5,.25) circle (2pt);
    \fill[\obscolor] (4.5,-1.25) circle (2pt);
\end{tikzpicture}}
\quad \raisebox{1cm}{\text{,}} \quad
 \begin{tikzpicture}
    \draw[thick, \bracolor] (0,0) ellipse (1 and .25);
    \draw[thick, \bradiffcolor] (0,-1) ellipse (1 and .25);

    \draw[thick ] (-1,0) to[out = 90, in = 180] (2.25,1.25) to[out = 0, in = 90](5.5,0) ;
    \draw[thick ] (1,0) to[out = 90, in = 180] (2.25,.5) to[out = 0, in = 90](3.5,0) ;

    \draw[thick ,\obscolor] (0,.25) to[out = 90, in = 180] (2.25,1) to[out = 0, in = 90](4.5,.25) ;
    
    \draw[thick,\bradiffcolor] (0+4.5,0) ellipse (1 and .25);
    \draw[thick, \bracolor] (0+4.5,-1) ellipse (1 and .25);

    \draw[thick ] (-1,-1) to[out = -90, in = 180] (2.25,-2.25) to[out = 0, in = -90](5.5,-1) ;
    \draw[thick ]  (1,-1) to[out = -90, in = 180] (2.25,-1.5) to[out = 0, in = -90](3.5,-1) ;
    \draw[thick ,\obscolor] (0,-1.25) to[out = -90, in = 180] (2.25,-2) to[out = 0, in = -90](4.5,-1.25) ;


    \fill[\obscolor] (0,.25) circle (2pt);
    \fill[\obscolor] (0,-1.25) circle (2pt);
    \fill[\obscolor] (4.5,.25) circle (2pt);
    \fill[\obscolor] (4.5,-1.25) circle (2pt);
\end{tikzpicture}
\end{align}
are disallowed by the gravitational path integral rules for the observer in \cite{lucaobserver, Akers:2025ahe} and are exponentially suppressed in the entropy of the observer in the rules of \cite{Harlow:2025pvj}.
Indeed, without these rules, the gravitational path integral appears to imply that the Hilbert space of closed universes is one-dimensional \cite{Almheiri:2019hni,Penington:2019kki,Marolf:2020xie,Usatyuk:2024mzs}. In that case, the no-boundary state is a pure state and is trivially the identity matrix.

Now suppose that we compute the gravitational inner product between two states specified by boundary data $\phi'$ and $\phi''$. In gravity, one computes inner products by filling in the region between the boundary conditions with all possible spacetimes, including higher topologies. This is again the same calculation one would do to compute matrix elements of the observer's no-boundary state
\begin{equation}
\bra{\phi''}\rho_\text{NB}\ket{\phi'}=~~~\mathrel{\raisebox{-0.5\height}{\begin{tikzpicture}
    \draw[thick, \bracolor] (0,0) ellipse (1 and .25);
    \draw[thick, \bracolor] (0,-1) ellipse (1 and .25);
    \draw[thick ] (-1,0) to[out = 90, in = 90] (-1.5,-.5) to[out = -90, in = -90](-1,-1) ;
    \draw[thick ] (1,0) to[out = 90, in = 90] (-2,-.5) to[out = -90, in = -90](1,-1) ;
    \draw[thick ,\obscolor] (0,.25) to[out = 90, in = 90] (-1.75,-.5) to[out = -90, in = -90](0,-1.25) ;

    \draw[thick, \bracolor] (0+4.5,0) ellipse (1 and .25);
    \draw[thick, \bracolor] (0+4.5,-1) ellipse (1 and .25);
    \draw[thick ] (-1+4.5,0) to[out = 90, in = 90] (-1.5+4.5,-.5) to[out = -90, in = -90](-1+4.5,-1) ;
    \draw[thick ] (1+4.5,0) to[out = 90, in = 90] (-2+4,-.5) to[out = -90, in = -90](1+4.5,-1) ;
    \draw[thick ,\obscolor] (0+4.5,.25) to[out = 90, in = 90] (-1.75+4.5,-.5) to[out = -90, in = -90](0+4.5,-1.25) ;

    \draw[thick] (2.25,-.5) to[out = 20, in = 160] (2.5,-.5)  ;
    
    \draw[thick] (2.15,-.5) to[out = -20, in = -160] (2.6,-.5)  ;
    \node[] at (1.5,-.5) {$+$};
    \node[] at (6.25,-.5) {$+\quad\dots$};
    \fill[\obscolor] (0,.25) circle (2pt);
    \fill[\obscolor] (0,-1.25) circle (2pt);
    \fill[\obscolor] (4.5,.25) circle (2pt);
    \fill[\obscolor] (4.5,-1.25) circle (2pt);
\end{tikzpicture}}}
\end{equation}
Therefore $\rho_\text{NB}=\mathbb{1}$ even after including topology change. It is important that when including topology change, the sum over geometries remains over spacetimes without boundary. 

We have focused on the Hilbert space of a single closed universe. This is because the observer only lives in a single universe. One may attempt to include baby universes in the path integral, but following the same logic as \cite{Almheiri:2019hni,Penington:2019kki,Marolf:2020xie,Usatyuk:2024mzs}, one will find that including these does not change the dimension of the Hilbert space, which is dictated only by the universe with the observer. 
In this case, time evolution also generates topology change. Therefore, even data $\phi'$ and $\phi''$ on spatial manifolds $\mathcal{M}'$ and $\mathcal{M}''$ with different topologies can be gauge equivalent.

When including topology change, quantum gravity is distinct from topological quantum field theory because in topological field theory, one does not naturally sum over topologies.\footnote{However, see \cite{Banerjee:2022pmw} for a study of what occurs when one does include a sum of topologies.} More precisely, in topological field theory one does not sum over geometries to compute an inner product between states, and even if one did, one would not allow topologies to connect different copies of an inner product calculation as one does in gravity when determining the dimensional of the Hilbert space via the replica trick.

\section{Closed universes in JT gravity}\label{sect:4.JT}
In section \ref{sect:2.obscomplete}, we argued on general grounds that the observer's no-boundary density matrix is the identity operator. As a consequence of their generality, the arguments were necessarily formal. Here, we complement that discussion with explicit checks in JT gravity with $\Lambda < 0$. Previous work on algebras in JT gravity with $\Lambda < 0$ has focused on open universes \cite{Penington:2023dql,Kolchmeyer:2023gwa, Penington:2024sum}. In JT gravity,
the Euclidean path integral that computes the no-boundary density matrix \eqref{eq:nb_density_filled} can be done exactly. We will match these Euclidean wormhole amplitudes \cite{sss,philsolo,disecting,lucavolume} with the matrix elements of the identity operator $\mathbb{1}$ on the physical Hilbert space. As in section \ref{sect:2.obscomplete}, we first consider no topology change. Topology change will be included in section \ref{subsect:JT4nonpert}.

\begin{figure}
 \centering
 \begin{tikzpicture}
    \draw[thick,\bradiffcolor] (0,0) ellipse (1 and .25);
    \draw[thick] (0,-1.5) to[out = 130, in = -90] (-1,0) to[out = 90, in = -130] (0,1.5) ;
        \draw[thick] (0,-1.5) to[out = 50, in = -90] (1,0) to[out = 90, in = -50] (0,1.5) ;
    \node at (1.25,0) {$b$};
     \draw[\obscolor,thick] (0,-1.5) to[out = 50, in = -90] (.5,0) to[out = 90, in = -50] (0,1.5) ;
         \fill[\obscolor] (0,-1.5) circle (2pt);
    \fill[\obscolor] (0,1.5) circle (2pt);
\end{tikzpicture}
 \caption{The Lorentzian spacetime with metric \eqref{3.9metric} with big bang and big crunch singularities. The time-reflection symmetric surface has the maximal length, $b$.}
 \label{fig:perogi}
\end{figure}
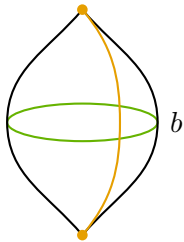
JT gravity is a theory of two-dimensional dilaton gravity with action
\begin{equation}
 I = \frac{1}{2}\int \d x \sqrt{g}\, \Phi (R+2)+\text{boundary terms}\,.
\end{equation}
The classical Lorentzian metric solutions are collapsing Milne universes with singularities at $t=\pm \pi/2$
\begin{equation}
 \d s^2=b^2\cos(t)^2\,\d x^2-\d t^2\,,\quad -\frac{\pi}{2}<t<\frac{\pi}{2}\quad 
 \label{3.9metric}
\end{equation}
as shown in figure \ref{fig:perogi}.
We consider
adding a quantum mechanical observer to this spacetime. This observer is perceived in the gravitational theory as a minimally coupled matter particle with energy $q$. 
For clarity's sake, both in our discussion of the Hilbert space and of the gravitational path integral, we first consider JT gravity without this observer particle.

\subsection{Hilbert spaces}\label{sect:3.1HJT}
We first review the Hilbert space of closed universes in JT gravity with $\Lambda < 0$, which was discussed in detail in \cite{Held:2024rmg}.\footnote{Similar discussion for more general dilaton potentials (including periodic ones) can be found in \cite{Blommaert:2025avl}.}$^,$\footnote{A version of this discussion actually also works for open universes (intervals). The WDW Hamiltonian constraint in that case generates a timelike flow in the patch, and is modified by boundary terms. Details of such a construction will be presented elsewhere.} In ADM variables, the most general metric configuration is
\begin{equation}\label{eq:JT_ADM}
 ds^2 = -N^2 \d t^2 + a^2 (\d x+N_\perp \d t)^2\,,\quad x\sim x+1\,.
\end{equation}
Following section \ref{sec:canonical_quant}, integrating $N$ and $N_\perp$ over the real line imposes the local Hamiltonian and momentum constraints, which are the generators of timelike and, respectively, spacelike diffeomorphisms. The canonical conjugates of $a$ and $\Phi$ are $p$ and $k$,
\begin{align}
 [a,p]=\i, \quad [\Phi,k]=\i.
\end{align}
We can gauge-fix $k$ and $a$ to be $x$-independent. The constraints then fix $\Phi$ and $p$ to also be $x$-independent \cite{Held:2024rmg}. This leaves us with a four-dimensional phase space of $x$-independent $(a,\Phi,p,k)$ that exactly describes JT gravity \cite{henneaux1985quantum}. In this description, all that remains from the constraints is the global Hamiltonian constraint
\begin{equation}
 H_\text{WDW}=-p k -a \Phi=0\,.\label{3.4con}
\end{equation}
The space of boundary conditions is two-dimensional. For instance, we can fix the length $a$ of the slice and the dilaton $\Phi$. As discussed in section \ref{sect:2.2} and \ref{sec:canonical_quant}, this auxiliary Hilbert space is larger than the physical Hilbert space. The latter is spanned by solutions of the Wheeler DeWitt equation, $H_{\text{WDW}}\ket{\psi} = 0$, which may be found by diagonalizing an operator which commutes with $H_\text{WDW}$, for instance
\begin{equation}\label{eq:b_def}
 b=\sqrt{a^2+k^2}\,.
\end{equation}
This has the geometric meaning of the maximal size of the universe \eqref{3.9metric}. Indeed, the classical solutions are $a=b\cos(t)$ and $k=b\sin(t)$. In this parameterization, one obtains
\begin{equation}
 H_\text{WDW}=-\i \frac{\d}{\d t}\,,
\end{equation}
so the WDW constraint generates timelike diffeomorphisms. Data that corresponds to different time slices in the same geometry describe the same quantum state. Thus, we can label physical states by $b$.
The wavefunctions that diagonalize $b$ and $H_\text{WDW}$ are 
\begin{equation}\label{eq:wf_L}
 \psi_b(a,\Phi)
 = \frac{1}{\sqrt{2\pi}}\frac{1}{\sqrt{b^2-a^2}}e^{\i \Phi \sqrt{b^2-a^2}}\, \equiv \langle a,\Phi | b\rangle .
\end{equation}
In the range of Euclidean wormholes solutions $t=\i \rho$ where $a>b$ this becomes
\begin{equation}\label{eq:wf_E}
 \psi_b(a,\Phi)
 =\frac{1}{\sqrt{2\pi}}\frac{1}{\sqrt{a^2-b^2}}e^{- \Phi \sqrt{a^2-b^2}}\,.
\end{equation}
This will be used in section \ref{subsect:JT1empty}. The inner product of physical states is computed using the expectation value of the rigging map \eqref{2.22rhoisdelta}
\begin{equation}
 \eta=\frac{1}{2\pi}\int_{-\infty}^{+\infty}\d N\,e^{-\i H_\text{WDW} N }\,.\label{eta}
\end{equation}
Alternatively, one can do a Faddeev–Popov gauge-fixing which results in the usual Klein-Gordon inner product on superspace \cite{Witten:2022xxp}. A convenient gauge-fixing for this calculation is to consider $a=0$ \cite{Blommaert:2025avl}. This immediately gives
\begin{equation}
\begin{split}
 \braket{b_1|b_2} & = -\frac{\i}{2}\int_{-\infty}^{+\infty} \d\Phi \bigg( \psi_{b_1}^*\frac{\d}{\d \Phi}\psi_{b_2}-\psi_{b_2}\frac{\d}{\d \Phi}\psi_{b_1}\bigg)
 = \frac{1}{b_1} \delta(b_1-b_2).\label{3.10ip}
\end{split}
\end{equation}
The identity matrix on the physical Hilbert space is therefore
\begin{align}
\label{eq:completenesse}
 \mathbb{1} = \int \d b\, b \ket{b}\bra{b}\,.
\end{align}

We now introduce one matter particle which represents the observer. Useful background and related comments can be found in \cite{Held:2024rmg,Penington:2023dql,Kolchmeyer:2023gwa,Kolchmeyer:2024fly}. Matter does not couple to the dilaton $\Phi$, so the classical solutions
are unaffected. We continue working in the same gauge for the geometric data. The location of the matter particle (observer) on the Cauchy slice is determined by an angular coordinate $\tau \sim \tau+b$ with $x=b \tau$. Before imposing the constraints, the phase space consists of $a,\Phi, \tau,q$ and their conjugates. One might now be tempted to claim that the Hilbert space consists of states $\ket{b}$ describing the spacetime, tensored with the Hilbert space of a matter particle on each fixed geometry. This would indeed be the case for open Cauchy slices in JT gravity describing eternal black holes \cite{Penington:2023dql,Kolchmeyer:2023gwa}. The states would then be labeled $\ket{b,\tau,q}$, or upon Fourier transforming $\ket{b,k,q}$.\footnote{For fixed $b$ and $q$ these span a complete basis for 
solutions of the AdS$_2$ Klein-Gordon equation, analogous to (3.10) in \cite{Kolchmeyer:2024fly} for dS$_2$.} However, we should recall that spatial translations 
are a gauge symmetry in gravity. Therefore, phase space is also subject to a momentum constrain, which in the presence of a matter particle reads \cite{Held:2024rmg}
\begin{equation}
 P_\text{matter}=k=0\,,\label{3.12}
\end{equation}
so the physical Hilbert space for a single particle is spanned by the states $\ket{b,k=0,q}$.\footnote{We note that in \cite{Iliesiu:2024cnh} this constraint seems to have been ignored.} The resolution of the identity is
\begin{align}
\label{eq:completenesOBS}
 \mathbb{1} = \sum_q\int \d b\, b \ket{b,k=0,q}\bra{b,k=0,q}\,.
\end{align}
It would be very interesting to give this observer access to matter quantum field theory operators along their worldline, but this goes beyond our present goals.

\subsection{The no-boundary state}\label{subsect:JT1empty}
A static observer in the spacetime \eqref{3.9metric}
has complete causal access. In section \ref{sect:2.obscomplete}, we explained that in such a scenario the observer's no-boundary state equals the identity on the physical Hilbert space
\begin{equation}
 \rho_\text{NB}=\mathbb{1}.
\end{equation}
We will now compute matrix elements of the no-boundary state in JT gravity using the Euclidean path integral and show that they are consistent with the Hilbert space discussion of section \ref{sect:3.1HJT}, supporting the general claim of section \ref{sect:2.obscomplete}.

For clarity, we will first consider the no-boundary state without an observer. However, to facilitate comparison with the observer's no-boundary state, we will still consider the cylinder topology instead of the disc
\begin{equation}
\bra{a_1,\Phi_1}\rho_{\text{NB}}\ket{a_2,\Phi_2} = \raisebox{-0.5\height}{\begin{tikzpicture}
    \draw[thick,\bracolor] (0,0) ellipse (1 and .2);
    \node at (1.7,0.05) {$a_1,\Phi_1$};
    \draw[thick, \bradiffcolor] (0,-1) ellipse (.5 and .125);
    \node at (.8,-1) {$b$};
    \draw[thick,\bracolor] (0,-2) ellipse (1.3 and .2);
    \node at (2,-2) {$a_2,\Phi_2$};
    \draw[thick] (-1,0) to[out = -60, in = 90] (-.5,-1)
    to[out = -90, in = 50] (-1.3,-2);
    \draw[thick] (1,0) to[out = 180+60, in = 90] (.5,-1)
    to[out=-90, in=180-50] (1.3,-2);
\end{tikzpicture}}
\end{equation}
We will comment on the Lorentzian calculation towards the end of this section. The path integral of JT gravity on Euclidean wormholes can be computed following \cite{sss} by separating it into two ``trumpets'' that are connected at the geodesic $b$ and including the appropriate Weil-Peterson measure
\begin{equation}
 \bra{a_1,\Phi_1}\rho_{\text{NB}}\ket{a_2,\Phi_2}=\int_0^\infty \d b\,b\,Z_\text{trumpet}(b,a_1,\Phi_1)Z_\text{trumpet}(b,a_2,\Phi_2)\,.\label{aphirhoaphi}
\end{equation}
The
trumpet partition functions with finite $a$ and $\Phi$ can be computed by 
a certain deformation of the Schwarzian path integral \cite{Iliesiu:2020zld}
\begin{equation}
 \raisebox{-0.5\height}{\begin{tikzpicture}
    \draw[thick,\bracolor] (0,0) ellipse (1 and .2);
    \node at (1.5,0) {$a,\Phi$};
    \draw[thick, \bradiffcolor] (0,-1) ellipse (.5 and .125);
    \node at (.7,-1) {$b$};
    \draw[thick] (-1,0) to[out = -60, in = 90] (-.5,-1);
    \draw[thick] (1,0) to[out = 180+60, in = 90] (.5,-1);
\end{tikzpicture}} = Z_\text{trumpet}(b,a,\Phi)\,.
\end{equation}
The calculation is greatly simplified by considering a subclass of boundary conditions corresponding with asymptotically Euclidean AdS
\cite{Maldacena:2016upp,Engelsoy:2016xyb,Jensen:2016pah}, 
\begin{equation}\label{eq:ass_bc}
 a = \frac{\beta}{\varepsilon}, \quad \Phi = \frac{1}{2\varepsilon}\,,\quad \varepsilon\to 0\,.
\end{equation}
Upon renormalization in the $\varepsilon\rightarrow0$ limit, both the Euclidean wavefunction \eqref{eq:wf_E} and the trumpet partition function 
become \cite{sss,Mertens:2019tcm}
\begin{equation}\label{eq:trumpet}
 Z_\text{trumpet}(b,\beta)=\psi_b(\beta) = \frac{1}{\sqrt{2\pi\beta}}e^{-b^2/4\beta}.
\end{equation}
Therefore, asymptotic matrix elements of the no-boundary state \eqref{aphirhoaphi} can be written as
\begin{equation}\label{eq:NB_id_JT}
 \bra{\beta_1}\rho_{\text{NB}}\ket{\beta_2}=\int_0^\infty\d b\, b\,\psi_b(\beta_1)\psi_b(\beta_2)=\bra{\beta_1}\mathbb{1}\ket{\beta_2}\,,
\end{equation}
so the wormhole amplitude is indeed the identity operator on the physical Hilbert space.
The integration over $b$ in \eqref{eq:completenesse} is a simple example of \eqref{eq:nb_density_squared}, with each of the trumpets computing a gauge-fixed matrix element of one no-boundary density matrix.

The matching \eqref{eq:NB_id_JT} can readily be extended to finite values of $a$ and $\Phi$. This is because the gravitational path integral has to satisfy the WDW equation, and so it necessarily decomposes into the basis of solutions \eqref{eq:wf_E}
\begin{equation}\label{eq:NB_a12}
 \bra{a_1,\Phi_1}\rho_{\text{NB}}\ket{a_2,\Phi_2}=\int_0^\infty \d b_1\int_0^\infty \d b_2\,\rho_\text{NB}(b_1,b_2)\,\psi_{b_1}(a_1,\Phi_1)\psi_{b_2}(a_2,\Phi_2)\,,
\end{equation}
with only the expansion coefficient $\rho_\text{NB}(b_1,b_2)$ left to be determined. Investigating this expression with asymptotic boundary conditions \eqref{eq:ass_bc} is sufficient to uniquely determine this expansion coefficient
\begin{equation}
 \rho_\text{NB}(b_1,b_2)=\frac{1}{b_1}\delta(b_1-b_2)\,.
\end{equation}
Similar logic was used to determine the finite cutoff disk amplitude in \cite{Iliesiu:2020zld}.

We pause to make a comment on the lapse contour.
 Consider a Lorentzian path integral with boundary conditions $(\ell_1,k_1)$ and $(\ell_2,k_2)$. The classical solutions \eqref{3.9metric} are $\ell=b\cos(t),\,k=b\sin(t)$. In quantum gravity, we must sum over all geometries connecting these boundary conditions. Classically, there are two contributions. Viewing $(\ell_1,k_1)$ as initial boundary condition, the classical solution is \eqref{3.9metric} evolved for a time $t_2-t_1$. The second contribution comes from viewing $(\ell_2,k_2)$ as the initial condition, in which case the classical solution is \eqref{3.9metric} evolved for a time $t_1-t_2$. If bras and kets are indistinguishable,\footnote{This seems consistent with \cite{Harlow:2023hjb} but potentially in tension with \cite{Witten:2025ayw}.} both $\braket{\ell_1,k_1|\ell_2,k_2}$ and $\braket{\ell_2,k_2|\ell_1,k_1}$ are non-zero, indicating that we count solutions with elapsed time $N>0$, and with $N<0$. So, the lapse $N$ is to be integrated from $-\infty$ to $+\infty$ in the Lorentzian path integral, which is consistent with section \ref{sect:2.obscomplete} and \cite{DiazDorronsoro:2017hti,Banihashemi:2024aal}. 

\subsection{The observer's no-boundary state}\label{sect:3.3ONBJT}
We next consider the Euclidean path integral with a single particle that represents an observer. In this case, the finite cutoff amplitude is not known, so we only consider asymptotic boundary conditions \eqref{eq:ass_bc}
\begin{equation}
\bra{\beta_1,\tau_1,q_1}\rho_\text{NB}\ket{\beta_2,\tau_2,q_2} = \raisebox{-0.5\height}{\begin{tikzpicture}
    \draw[thick,\bracolor] (0,0) ellipse (1 and .2);
    \node at (2,0.05) {$\beta_1,\tau_1,q_1$};
    \draw[thick, \bradiffcolor] (0,-1) ellipse (.5 and .125);
    \node at (.8,-1) {$b$};
    \draw[thick,\bracolor] (0,-2) ellipse (1.3 and .2);
    \node at (2.35,-2) {$\beta_2,\tau_2,q_2$};
    \draw[thick] (-1,0) to[out = -60, in = 90] (-.5,-1)
    to[out = -90, in = 50] (-1.3,-2);
    \draw[thick] (1,0) to[out = 180+60, in = 90] (.5,-1)
    to[out=-90, in=180-50] (1.3,-2);
    \draw[thick, \obscolor] (.5,-.15) to[out=-100,in=40] (-.5,-2.15);
    \fill[\obscolor] (.5,-.15) circle (2pt);
    \fill[\obscolor] (-.5,-2.15) circle (2pt);
\end{tikzpicture}}.
\end{equation}
This path integral was computed in \cite{Blommaert:2018iqz,philsolo}. We will find it convenient to present the result by an inverse Laplace transform to fixed energy holographic boundary conditions
\begin{equation}
 \ket{\beta}=\int_0^\infty \d E\,\rho(E)\,e^{-\beta E}\ket{E}\,,
\end{equation}
with $\rho(E)$ the Schwarzian spectral density \cite{Stanford:2017thb}. This leads to\cite{Yang:2018gdb,Iliesiu:2024cnh}
\begin{equation}\label{eq:NB_E12}
 \bra{E_1,\tau_1,q_1}\rho_\text{NB}\ket{E_2,\tau_2,q_2} = \frac{\Gamma(\Delta)^2\Gamma(\Delta\pm 2\i\sqrt{E_1})}{\Gamma(2\Delta)}\frac{1}{\rho(E_1)}\delta(E_1-E_2)\,\delta_{q_1 q_2}\, ,\quad \Delta(\Delta-1) = q_1^2\ ,
\end{equation}
where the $\pm$ signify a product over $\Gamma$ functions with both signs.

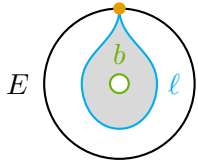
\begin{figure}[t]
 \centering
 \begin{tikzpicture}
    \draw[thick] (0,0) circle (1);
    
    \filldraw[gray, opacity = .3] (0,1)to[out = -90, in = 90](-.5,0) to[out= -90, in =180](0,-.6) to[out = 0,in =-90] (.5,0)to[out = 90,in = -90] (0,1) ;
    \draw[thick,cyan] (0,1)to[out = -90, in = 90](-.5,0) to[out= -90, in =180](0,-.6) to[out = 0,in =-90] (.5,0)to[out = 90,in = -90] (0,1) ;
    \filldraw[white] (0,0) circle (.125);
    \draw[thick, \bradiffcolor] (0,0) circle (.125);
    \node[\bradiffcolor] at (0,.4) {$b$};
    \node[] at (-1.35,0) {$E$};
    \node[cyan] at (.73,0) {$\ell$};
    \filldraw[\obscolor] (0,1) circle (.075);
\end{tikzpicture}
 \caption{Open geodesic $\ell$ (cyan) which ends at the insertion of the observer particle near the asymptotic boundary (orange) in a spacetime with fixed energy asymptotic boundary conditions. This geodesic is homologous to a closed geodesic of length $b$ (green). The amplitude for the white and gray regions are \eqref{3.29psieell} and \eqref{3.36psibell} respectively.}
 \label{fig:teardrop_tikz}
\end{figure}

We would like to show that \eqref{eq:NB_E12} is consistent with interpreting the wormhole as an identity operator 
\eqref{eq:completenesOBS}.\footnote{It is difficult to directly solve the WDW equation for this configuration because of the point-like nature of the observer. Our goal in this section is more modest than in the previous one. We demonstrate consistency by using a hybrid between the path integral and canonical arguments. In appendix \ref{app:cosmo}, we consider a dynamical cosmological constant as some simpler model for an observer. In that case, one can completely separate the path integral and canonical arguments and find agreement.}
It is useful to rewrite the identity operator with an explicit momentum constraint
\begin{equation}\label{3.27one}
\begin{split}
 \mathbb{1}&=\int_0^\infty \d b\,b\ket{b,k=0}\bra{b,k=0}
 =\int_0^\infty \d b\,b\sum_{k=-\infty}^{+\infty}
 \eta_\text{momentum} \ \ket{b,k}\bra{b,k}
 \,,
\end{split}
\end{equation}
with the momentum rigging map
\begin{equation}
 \eta_\text{momentum}=\frac{1}{b}\int_0^b \d N_\perp\,e^{\i P_\text{matter} N_\perp}.\label{etamom}
\end{equation}
In equation \eqref{3.27one}, we suppressed the $q$ label, and will continue to do so throughout this section to avoid clutter. To compute the matrix element of \eqref{3.27one} in asymptotic states $\ket{E,\tau}$, a convenient intermediate step is to transform from fixed energy $E$ states to fixed asymptotic geodesic length $\ell$ states (see figure \ref{fig:teardrop_tikz}). The wavefunctions are \cite{Blommaert:2018oro,Yang:2018gdb}
\begin{equation}
 \ket{E}=\int_{-\infty}^{+\infty}\d \ell \psi_E(\ell)\ket{\ell}\,,\quad \psi_E(\ell)=K_{2\i E^{1/2}}(e^{-\ell/2})\,,\label{3.29psieell}
\end{equation}
where $K$ is a Bessel function.
The matrix elements of the identity operator \eqref{3.27one} in this basis then decompose as
\begin{align}
 \bra{\ell_1,\tau_1}\mathbb{1}&\ket{\ell_2,\tau_2}=\int_0^\infty\d b \sum_{k=-\infty}^{+\infty}\int_{0}^{b}\d \tau\, e^{\i k \tau}\braket{\ell_1,\tau_1|b,k}\braket{b,k|\ell_2,\tau_2}\label{3.20}\\&=\int_0^\infty\d b \int_{0}^{b}\d \tau\sum_{k=-\infty}^{+\infty}\braket{\ell_1,\tau_1|b,k}\braket{b,k|\ell_2,\tau_2+\tau}=\int_0^\infty \d b \braket{\ell_1|b}\braket{b|\ell_2}\int_0^b\d \tau \braket{\tau_1|\tau_2+\tau}_b\,.\nonumber
\end{align}
Here, $\braket{\tau_1|\tau_2}_b$ is the inner product between matter states obtained by inserting matter particles in the geometry \eqref{3.9metric} with size $b$. By the definition of $\ket{\ell,\tau}$, the matter states are prepared by inserting a matter particle of mass $q$ at the asymptotic Euclidean boundary at spatial coordinate $\tau$, while fixing the length $\ell$ of the geodesic that is homologous to $b$ and ending on the location of this operator insertion (see figure \ref{fig:teardrop_tikz}).
The matter propagator in this geometry equals\footnote{To compute this, one takes the asymptotic limit of the bulk-bulk correlator in AdS$_2$ with the sum over windings. In the asymptotic limit this becomes just $e^{-\Delta d(\tau_1,\tau_2)}$.} 
\begin{equation}
 \braket{\tau_1|\tau_2}_b=G_\Delta(\tau_1,\tau_2)=e^{-(\ell_1+\ell_2)\Delta/2}\sinh(b/2)^{2\Delta}\sum_{w=-\infty}^{+\infty}\frac{1}{\cosh((\tau_2-\tau_1+b w)/2)^{2\Delta}}\,.\label{3.18tautau}
\end{equation}
The sum over $w$ corresponds with the sum of windings of the particle around the spatial circle
\begin{equation}
\braket{\tau_1|\tau_2}=\raisebox{-0.5\height}{\begin{tikzpicture}
    \draw[thick,\bracolor] (0,0) ellipse (1 and .2);
    \node at (1.5,0) {$\ell_1$};
    \draw[thick,\bracolor] (0,-2) ellipse (1.3 and .2);
    \node at (1.8,-2) {$\ell_2$};
    \draw[thick] (-1,0) to[out = -60, in = 90] (-.5,-1)
    to[out = -90, in = 50] (-1.3,-2);
    \draw[thick] (1,0) to[out = 180+60, in = 90] (.5,-1)
    to[out=-90, in=180-50] (1.3,-2);
    \draw[thick, \obscolor] (.5,-.15) to[out=-100,in=40] (-.5,-2.15);
    \fill[\obscolor] (.5,-.15) circle (2pt) node[below left,black]{$\tau_1$}
    -- 
    (-.5,-2.15) circle (2pt) node[below,black]{$\tau_2$};
\end{tikzpicture}} + \raisebox{-0.5\height}{\begin{tikzpicture}
    \draw[thick,\bracolor] (0,0) ellipse (1 and .2);
    \node at (1.5,0) {$\ell_1$};
    \draw[thick,\bracolor] (0,-2) ellipse (1.3 and .2);
    \node at (1.8,-2) {$\ell_2$};
    \draw[thick] (-1,0) to[out = -60, in = 90] (-.5,-1)
    to[out = -90, in = 50] (-1.3,-2);
    \draw[thick] (1,0) to[out = 180+60, in = 90] (.5,-1)
    to[out=-90, in=180-50] (1.3,-2);
    
    \draw[thick, \obscolor] (.5,-.15) to[out=-100,in=20] (-.5,-1);
    \draw[thick, dashed, \obscolor] (-.5,-1) to[out=-30,in=180] (.75,-1.5);
    \draw[thick, \obscolor] (.75,-1.5) to[out=220,in=40] (-.5,-2.15);    
    \fill[\obscolor] (.5,-.15) circle (2pt) node[below left,black]{$\tau_1$}
    -- 
    (-.5,-2.15) circle (2pt) node[below,black]{$\tau_2$};
\end{tikzpicture}} + \quad\dots\, .
\end{equation}
Note that the geometry is completely fixed once we specify $\ell_1$, $\ell_2$, $b$, $\tau_1$, and $\tau_2$. This allows computation of \eqref{3.18tautau}, using suitable holographic renormalization. Finally, to compute \eqref{3.20}, we need the gravitational transformation from the $b$ basis to the $\ell$ basis
which reads\footnote{The action for this piece of spacetime (gray region in figure \ref{fig:teardrop_tikz}) comes from the corner term in the JT action. Denoting the angle enclosed by the geodesic $\alpha$ one finds using hyperbolic geometry $\alpha=-2\varepsilon\, e^{-\ell/2}\cosh(b/2)$ with the action $\alpha\Phi=\alpha/2\varepsilon$.}
\begin{equation}
 \braket{\ell|b}=e^{-e^{-\ell/2}\cosh(b/2)}.\label{3.36psibell}
\end{equation}
Defining $\psi_b(\ell)=e^{-\ell \Delta/2}\sinh(b/2)^\Delta \braket{\ell|b}$, one can explicitly compute the matrix element of the identity operator \eqref{3.27one} in asymptotic fixed energy basis\footnote{The twist integral in \eqref{3.20} is extended to the full real axis by leveraging the sum over windings in \eqref{3.18tautau}. The integral over twists gives
\begin{equation}
 \int_{-\infty}^{+\infty}\d \tau \frac{1}{\cosh(\tau/2)^{2\Delta}}=\frac{4^\Delta \Gamma(\Delta)^2}{\Gamma(2\Delta)}\,.
\end{equation}
In addition, we explicitly evaluated the following integral using (6.624.3) \cite{gradshteyn2014table} and some further simplifications
\begin{equation}
 \psi_E(b)=\int_{-\infty}^{+\infty}\d\ell\, \psi_E(\ell)\psi_b(\ell)= {i^{\frac{1}{2}-\Delta} \sqrt{2\pi}}\Gamma(\Delta\pm 2\i \sqrt{E})\sqrt{\sinh(b/2)}P_{2\i \sqrt{E}-1/2}^{1/2-\Delta}(\cosh(b/2)).
\end{equation}
with $P_l^m$ the associated Legendre functions. Completeness follows from the
Mehler-Fock transform
\begin{equation}
 \int_0^\infty \d E\,\rho(E) \frac{1}{\Gamma(\Delta\pm 2\i\sqrt{E})}\psi_E(b_1)\psi_E^*(b_2)=\delta(b_1-b_2)\,.
\end{equation}
Their orthogonality relation was used to obtain the delta function in \eqref{3.27}. This slightly improves some equations in \cite{Iliesiu:2024cnh}. They consider what they call a basis $\ket{b,u}$ which we labeled $\ket{b,\tau}$. This is over-complete because the $\tau$ label is redundant. We can see this explicitly from these calculations but also from the general discussion on gauging spatial diffeomorphisms around \eqref{3.12}.}
\begin{equation}
\begin{split}
&\bra{E_1,\tau_1}\mathbb{1}\ket{E_2,\tau_2}\\
 &=\int_0^\infty\d b\int_{-\infty}^{+\infty}\d \ell_1 \psi_{E_1}(\ell_1)\psi_b(\ell_1)\int_{-\infty}^{+\infty}\d \ell_2 \psi_{E_2}(\ell_2)\psi_b(\ell_2)\int_{-\infty}^{+\infty}\d \tau \frac{1}{\cosh((\tau+\tau_2-\tau_1)/2)^{2\Delta}}\\
 &=
 {4^\Delta}
 \frac{\Gamma(\Delta)^2\Gamma(\Delta\pm 2\i\sqrt{E_1})}{\Gamma(2\Delta)}\frac{1}{\rho(E_1)}\delta(E_1-E_2)\,.\label{3.27}
\end{split}
\end{equation}
This matches with the evaluation of matrix elements of the no-boundary state \eqref{eq:NB_E12} using the Euclidean gravitational path integral which used different techniques \cite{Blommaert:2018iqz,philsolo}. 

A takeaway point of this calculation is that we learned that the Euclidean gravitation path integral implements naturally the projector $\eta_\text{momentum}$ \eqref{etamom} that imposes the momentum constraint, in addition to the Hamiltonian constraint \eqref{eta}. This was not manifest in earlier calculations of this wormhole amplitude.
The no-boundary state is thus the tracial state on the physical Hilbert space
\begin{align}
 \Tr ( \cdot )=\int_0^\infty\d b\,b \bra{b,k=0}\cdot \ket{b,k=0}\,.
\end{align}
Of course, one should consider reasonable operators for which this converges, which does not include e.g.~simple functions of $b$. 

The above construction generalizes to the Hilbert space of closed universes in JT with $n$ matter particles, which is spanned by states of the form
\begin{equation}
 \ket{b,k_1\dots k_n},\quad \sum_{i=1}^n k_i=0.
\end{equation}
The matter Fock space is constructed on spacetimes \eqref{3.9metric} specified by the geodesic length $b$. The trace on the sector with $n$ identical particles is
\begin{equation}
 \Tr(\cdot)=\int_{0}^\infty \d b\,\frac{1}{n!}\sum_{k_1=-\infty}^{+\infty}\dots \sum_{k_n=-\infty}^{+\infty}\int_0^b\d N_\perp\, e^{\i N_\perp (k_1+\dots +k_n)} \bra{b_1,k_1\dots k_n}\cdot \ket{b,k_1\dots k_n}\,.
\end{equation}
The wormhole with $n$ particles inserted on Euclidean asymptotic boundaries computes an inner product of the identity in this Hilbert space, but writing out this amplitude is messy for $n>1$.

\subsection{Including topology change}\label{subsect:JT4nonpert}
To conclude this section, 
examine the effect of topology change. To our knowledge, it is not understood what the WDW Hamiltonian on phase space is that generates topology change. Thus, we limit ourselves to interpreting the answer of the gravitational path integral as a suggestion for what the relevant Hilbert spaces are. The fact that we sum over all spacetimes consistent with boundary conditions shows that one can not distinguish $\rho_\text{NB}$ from $\rho_\text{NB}^2$.

We first consider, as in section \ref{subsect:JT1empty}, the case of JT gravity without observer. Summing over topologies in 
the gravitational path integral,
in JT gravity 
one finds \cite{sss}\footnote{This is computed by summing over all topologies connecting the two asymptotic boundaries
\begin{equation}
    Z(\beta_1.\beta_2)=\int_{-\infty}^{+\infty}\d E_1\,e^{-\beta_1 E_1}\int_{-\infty}^{+\infty}\d E_2\,e^{-\beta_1 E_2}\average{\rho(E_1)\rho(E_2)}\,.
\end{equation}
Here $\average{\rho(E_1)\rho(E_2)}$ decomposes into a genus expansion associated with the different gravitational topologies. This expansion matches exactly with the expansion of computing a certain observable in some particular random matrix integral \cite{sss}
\begin{equation}
    \average{\rho(E_1)\rho(E_2)}=\frac{1}{\mathcal{Z}}\int \d H\,e^{-V(H)}\Tr \delta(H-E_1)\Tr\delta(H-E_1)\,.
\end{equation}
In one member of the ensemble, which also has a gravity interpretation as some theory of dilaton gravity \cite{Blommaert:2021fob}, the integral over $H$ collapses to a fixed Hamiltonian $H_0$ with eigenvalues $E_i$ and one recovers the claimed equation.
}
\begin{equation}
 \bra{\beta_1}\rho_\text{NB}\ket{\beta_2}=\sum_{i=1}^N e^{-\beta_1 E_i}\sum_{j=1}^N e^{-\beta_2 E_j}\,.
\end{equation}
We imagine working in one member of the JT ensemble \cite{Blommaert:2019wfy,Marolf:2020xie,Blommaert:2021fob}, where $E_i$ are fixed eigenvalues. In this expression, there is no correlation between the bra and the ket $\ket{\beta}$. This admits an interpretation as inserting an identity operator in a ``physical Hilbert space'' if and only if that physical Hilbert space is one dimensional, $\mathcal{H} = \ket{E_1\dots E_N}$, where
\begin{equation}
 \braket{\beta_1|E_1\dots E_N}=\sum_{i=1}^N e^{-\beta_1 E_i}\,.
\end{equation}
The fact that the Hilbert space without the observer is one-dimensional remains true after we average over eigenvalues using the matrix integral, roughly speaking because the average of one is one \cite{Usatyuk:2024mzs}. With the observer one finds from explicit JT gravity calculations \cite{disecting,philsolo}
\begin{equation}
 \bra{\beta_1,\tau_1,q_1}\rho_\text{NB}\ket{\beta_2,\tau_2,q_2}=\sum_{i=1}^Ne^{-(\beta_1+\beta_2)E_i}\frac{\Gamma(\Delta)^2\Gamma(\Delta\pm 2\i \sqrt{E_i})}{\Gamma(2\Delta)}\,\delta_{q_1q_2}\,,\quad \Delta(\Delta-1)=q_1^2\,.
\end{equation}
The key feature is the diagonal sum over $E_i$ correlating the bra and ket. This admits an interpretation as inserting a density operator in a ``physical Hilbert space'' if and only if that physical Hilbert space is (for a given $q$) $N$ dimensional and spanned by orthogonal states $\{\ket{E_i} \}$. 
This is a weight because the algebra is type I$_{\infty}$ ($N = \infty$). For UV completions of JT gravity, such as sine dilaton gravity \cite{Blommaert:2025avl}, the algebra is finite-dimensional, type I$_{N}$, so the no-boundary state is a normalizable state.

\section{Observers with causal horizons}\label{sect:3obshorizon}
Thus far, we have considered density matrices for the entire universe. It is important to understand the more generic case where an observer experiences a horizon, the canonical example being de Sitter space. We will proceed in two complementary ways. First, we formalize the statement that, if the density matrix on the global slice is the identity, then the density matrix on a subalgebra is also the identity on that reduced algebra. We then work more constructively, providing a similar path integral approach to section \ref{sect:2.obscomplete}. This serves as a consistency check because our formal arguments must match known results in de Sitter space derived from a different direction \cite{Chandrasekaran:2022cip}.
\subsection{The algebraic approach}
In the perturbative $G\to 0$ limit, we can use semiclassical tools to study the no-boundary proposal. In this limit, we can consider a specific classical background $\mathcal{M}$ and study the perturbative (GNS) Hilbert space around it, which we denote $\mathcal{H}_\mathcal{M}$. 
In the previous sections, we discussed the global Hilbert space $\mathcal{H}$ involving all possible backgrounds. Here, we will assume the perturbative Hilbert space $\mathcal{H}_\mathcal{M}$ can be understood as a superselection sector of $\mathcal{H}$, as is familiar from the quantum error correction interpretation of AdS/CFT \cite{Almheiri:2014lwa, Harlow:2016vwg}. The Hilbert space description $\mathcal{H}_\mathcal{M}$ should hold for generic low energy excitations in the perturbative limit.

\begin{figure}
 \centering
 \begin{tikzpicture}[scale  = 1.25]
    \draw[thick] (-2,0)-- (0,2);
    \draw[thick] (0,0)--(-2,2);
    \draw[thick] (0,0)-- (2,2);
    \draw[thick] (2,0)--(0,2);
    \draw[thick] (2,0)-- (4,2);
    \draw[thick] (4,0)--(2,2);
    \draw[thick] (4,0)-- (6,2);
    \draw[thick] (6,0)--(4,2);
    \draw[thick] (6,0)-- (8,2);
    \draw[thick] (8,0)--(6,2);
    \draw[thick] (8,0)-- (10,2);
    \draw[thick] (10,0)--(8,2);
    \draw[double, thick ] (2,2) -- (4,2);
    \draw[double, thick ] (2,0) -- (4,0);
    \draw[double, thick ] (0,0) -- (-2,0);
    \draw[double, thick ] (0,2) -- (-2,2);
    \draw[double, thick ] (8,0) -- (6,0);
    \draw[double, thick ] (8,2) -- (6,2);
    \draw[decorate, decoration={zigzag}, thick] (0,2) -- (2,2);
    \draw[decorate, decoration={zigzag},thick] (0,0) -- (2,0);
    \draw[decorate, decoration={zigzag}, thick] (4,2) -- (6,2);
    \draw[decorate, decoration={zigzag},thick] (4,0) -- (6,0);
    \draw[decorate, decoration={zigzag}, thick] (8,2) -- (10,2);
    \draw[decorate, decoration={zigzag},thick] (8,0) -- (10,0);
    \draw[thick, \obscolor] (4,0) -- (4,2);
    \draw[] (10,0) -- (10,2);
    \draw[] (-2,0) -- (-2,2);
    \draw[] (-2.125,1.125)--(-2,1) -- (-1.875,1.125);
    \draw[] (-2.125+12,1.125)--(-2+12,1) -- (-1.875+12,1.125);
        \draw[] (-2.125,1.125-.125)--(-2,1-.125) -- (-1.875,1.125-.125);
    \draw[] (-2.125+12,1.125-.125)--(-2+12,1-.125) -- (-1.875+12,1.125-.125);
         \fill[\obscolor] (4,0) circle (2pt); 
     \fill[\obscolor] (4,2) circle (2pt); 
\end{tikzpicture}
 \caption{SdS with $n = 3$. An observer (orange) cannot distinguish different values of $n$. The arrows denote a periodic identification.}
 \label{fig:SdS_tikz}
\end{figure}
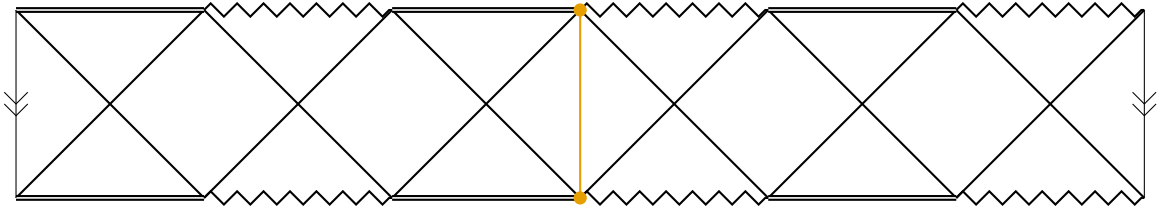

Within a superselection sector $\mathcal{H}_\mathcal{M}$ the observer has access to an algebra corresponding to its causal diamond, $\mathcal{A}_{obs}^{\mathcal{M}}$. $\mathcal{A}_{obs}^{\mathcal{M}}$ and its commutant on $\mathcal{H}_{\mathcal{M}}$ are subalgebras of a larger algebra of (bounded) operators $\mathcal{B}(\mathcal{H})$. There is some degeneracy in specifying the commutant, as globally distinct classical backgrounds can be indistinguishable in the observer's causal diamond. This is a source of entropy. One example is Schwarzschild de Sitter space (SdS). There are solutions that consist of $n$ black holes and $n$ cosmological horizons cyclically glued (see figure \ref{fig:SdS_tikz}).

The no-boundary path-integral 
can be used to define expectation values $\langle ... \rangle_\text{NB}$ for the observer's algebra that satisfies the properties of a trace on $\mathcal{B}(\mathcal{H})$
\begin{align}
\label{eq:traceprop}
 \braket{ab}_\text{NB} = \braket{ba}_\text{NB}\,,\quad \forall a,b \in \mathcal{A}_\text{obs}\,.
\end{align}
Let us now consider operators $a$ in $\mathcal{A}_{obs}^{\mathcal{M}}$. If the subalgebra $\mathcal{A}_{obs}^{\mathcal{M}}$ is type I or II, it has a trace $\text{Tr}(...)$. Assuming that $\mathcal{A}_{obs}^{\mathcal{M}}$ does not have a center, this trace is unique up to overall rescalings. Therefore, it must be related with $\rho_\text{NB}$ up to an overall constant\footnote{\label{footnote}If $\mathcal{A}_{obs}^{\mathcal{M}}$ is of type III, the expectation value of any operator with respect to the no-boundary state is infinite. Then \eqref{eq:traceprop} is a vacuous statement because both sides are divergent.}
\begin{equation}\label{eq:tr_Tr}
 \langle a \rangle_\text{NB} = \mathcal{Z}'_\mathcal{M} \cdot \text{Tr}(a),
\end{equation}
with 
\begin{equation} \label{eq:Z_prime}
 \mathcal{Z}'_\mathcal{M} = \frac{\mathcal{Z}_\mathcal{M}}{\text{Tr}(\mathbb{1}_{\mathcal{M}})},
\end{equation}
and $\mathcal{Z}_\mathcal{M} = \langle \mathbb{1}_{\mathcal{M}} \rangle_\text{NB}$. Recall that the definition of the no-boundary state is through the gravitational path integral. Therefore, $\mathcal{Z}_\mathcal{M}$ can be understood as the no-boundary path integral around the saddle $\mathcal{M}$. For example, if we take $\mathcal{M}$ to be the static patch of de Sitter space, then the no-boundary path integral is dominated by the sphere partition function. To leading order in $G$, one would find $\log \mathcal{Z}_\mathcal{M} = \frac{A}{4G}$ \cite{gibbons1977action}. In this case the algebra is type II$_1$, which enables us to canonically choose a rescaling such that $\text{Tr}(\mathbb{1}_{\mathcal{M}}) = 1$, leading to (ignoring the subleading corrections to $\mathcal{Z}_\mathcal{M}$)
\begin{equation}\label{eq:ds_trace}
 \langle a \rangle_\text{NB} = e^{\frac{A}{4G}} \cdot \text{Tr}(a).
\end{equation}
In a more general background however, \eqref{eq:Z_prime} is more subtle. The problem is that generically we expect both $\mathcal{Z}_\mathcal{M}$ and $\text{Tr}(\mathbb{1})$ to diverge. This happens both when the algebra is type I$_\infty$ or II$_\infty$, such as in SdS \cite{Kudler-Flam:2023qfl}, dS with a massless scalar field \cite{Kudler-Flam:2025pol}, or inflationary cosmology \cite{Kudler-Flam:2024psh, Chen:2024rpx, Speranza:2025joj}. However, we expect that in these cases a proper renormalization scheme renders $\mathcal{Z}'_\mathcal{M}$ finite. For example, one may regularize the algebra by taking a limit of projection operators in the trace, as is done in section 2 of \cite{Chen:2024rpx}.
Assuming such a scheme is well-defined and defining $\langle a \rangle_\text{NB} = \text{Tr}(\rho_\text{obs}^\mathcal{M}\,a)$, we find that 
\begin{align}
\rho_\text{obs}^\mathcal{M} = \mathcal{Z}'_{\mathcal{M}}\mathbb{1}_{\mathcal{M}}. 
\end{align}

A slightly more general setup with multiple dS vacua as superselection sectors was considered in \cite{witten2024background}, in which case (formally, as we are omitting the perturbative and non-perturbative corrections from each sector, and between the sectors)
\begin{align}
\label{eq:rhodsvacua}
 \rho_\text{NB} =\bigoplus_{\alpha} e^{\frac{A_{\alpha}}{4G}}\mathbb{1}_{\mathcal{M_{\alpha}}}.
\end{align}
When computing the von Neumann entropy of a state in one vacuum $\alpha$, one can appeal to the relative entropy of this normalized state with respect to the unnormalized no-boundary state $\rho_\text{NB}$
\begin{align}
 S_\text{vN}(\rho) = -S_\text{rel}(\rho\parallel \rho_\text{NB}) = -S_\text{rel}(\rho\parallel e^{\frac{A_{\alpha}}{4G}}\mathbb{1}_{\mathcal{M}_{\alpha}})= -S_\text{rel}(\rho\parallel \mathbb{1}_{\mathcal{M}_{\alpha}})+\frac{A_{\alpha}}{4G}\,.
\end{align}
The first term is the $O(1)$ entropy of \cite{Chandrasekaran:2022cip} and the second term is this $O(1/G)$ contribution that depends on the subsector. Thus, there is no additive ambiguity when interpreted as an asymptotic expansion. For a general density matrix (in the full Hilbert space $\mathcal{H}$), the entropy is given by
\begin{equation}
 S_\text{vN}(\rho) = -S_\text{rel}\left(\rho \parallel \rho_\text{NB}\right)
 = -\langle \rho \log \rho \rangle_\text{NB}.
\end{equation}
Here the normalization used is $\langle \rho \rangle_\text{NB}=1$.
Now consider a density matrix $\rho$ affiliated to $\mathcal{A}_{obs}^\mathcal{M}$. By \eqref{eq:tr_Tr} the same normalization would mean $\text{Tr}\rho = (\mathcal{Z}'_\mathcal{M})^{-1}$. Defining the normalized $\tilde \rho = \mathcal{Z}'_\mathcal{M} \cdot \rho$ (with $\text{Tr}\tilde \rho = 1$), we may use \eqref{eq:tr_Tr} to find
\begin{equation}
 \begin{split}
 S_\text{vN}(\rho) &= 
 \log \mathcal{Z}'_{\mathcal{M}}-S_\text{rel}(\tilde \rho\parallel \mathbb{1}_{\mathcal{M}})= \log \mathcal{Z}'_{\mathcal{M}}-\text{Tr}\left(\tilde\rho \log \tilde\rho\right).
 \label{actionentropy4.11}
 \end{split}
\end{equation}
Here $S_\text{rel}$ is defined using the perturbative trace.
If $\mathcal{Z}'_\mathcal{M}$ is dominated by a classical no-boundary saddle, we find to leading order in $G$ that the entropy is minus the on-shell action of that saddle
\begin{align}
 S_\text{vN}(\rho) = \log \mathcal{Z}'_\mathcal{M} +O(1) = -I_{\mathcal{M}} +O(1)\,.
\end{align}
Thus, the no-boundary state enables us to compute absolute entropy.

\subsection{A qubit example}
In order to gain further intuition for what is occuring in a completely controllable environment, it is useful to revisit the Araki-Woods construction of type II algebras \cite{araki1968classification}. Consider the Hilbert space of $2N$ qubits $\mathcal{H}_N = (\mathbb{C}^{2N})_R \otimes (\mathbb{C}^{2N})_L$, where the subscripts denote qubits on the ``right'' and the ``left.'' This is the analog of a single $\mathcal{H}_{\mathcal{M}}$ with $N$ playing the role of $1/G$. We will later generalize to the direct sum of several Hilbert spaces in analogy to $\mathcal{H}$ itself. The complete algebra of operators acting on both sides is type I by construction. 
We consider a state that consists of each qubit in the right maximally entangled with the corresponding qubit in the left in a Bell pair
\begin{align}
\label{eq4.12}
 \ket{\Psi_N} = \frac{1}{2^{N/2}}\bigotimes_i^N(\ket{00}_i + \ket{11}_i).
\end{align}
The reduced density matrix on the right is given by the identity operator
\begin{align}
 \rho_R = 2^{-N}\mathbb{1}_R.
\end{align}
As an operator on the joint Hilbert space $\mathcal{H}_N$, this is of course $\rho_R \otimes \mathbb{1}_L=\tilde{\rho}_R$.
This implies that $\ket{\Psi_N}$ is a tracial state for the algebra of operators acting nontrivially only on right side
\begin{align}
 \bra{\Psi_N}a_R b_R \ket{\Psi_N} = \bra{\Psi_N}b_R a_R \ket{\Psi_N} , \quad \forall a_R ,b_R \in \mathcal{A}_R.
\end{align}
Indeed, it is straightforward to check that this is simply the matrix trace on $(\mathbb{C}^{2N})_R$, normalized such that the trace of the identity is one
\begin{align}
 \bra{\Psi_N}a_R \ket{\Psi_N} =2^{-N}\Tr_R(a_R) = 2^{-N}\sum_{i = 1}^{N}\left(\bra{0}a_R\ket{0}_i + \bra{1}a_R\ket{1}_i \right).
\end{align}
In comparing to CLPW, $\ket{\Psi_N}$ is the analog of the global Bunch-Davies state with an observer.

Note that $\ket{\Psi_N}$ is not the unique way to extend $\rho_R$ to a state on the joint left-right system. For instance, the partial trace of the joint identity operator reduces to $\rho_R$
\begin{align}
 \Tr_L(2^{-2N}\mathbb{1}_R\otimes \mathbb{1}_L) = \rho_R.
\end{align}
The unnormalized version of this operator, $\mathbb{1}_R\otimes \mathbb{1}_L$, is analogous to the unnormalized no-boundary state on $\mathcal{M}$.
In the thermodynamic limit ($N\rightarrow \infty$), this state does not lead to finite expectation values for any element of $\mathcal{A}_R$ unless the normalization of $2^{-2N}$ is included. We can, however, consider it as a formal asymptotic expansion in $N$
\begin{align}
 \mathbb{1}_R\otimes \mathbb{1}_L=
 2^{N} \tilde{\rho}_R.
\end{align} 
The diverging prefactor $2^{N}$ is the analog of $\mathcal{Z}'_{\mathcal{M}}$ while the remaining operator is a valid state for operators on $\mathcal{A}_R$.
The reduced state $\tilde{\rho}_R$ is the analog of $\mathbb{1}_{\mathcal{M}}$. In this construction, $\Tr_R \rho_R = 1$, equivalently, the normalized trace of the identity is one, $\bra{\Psi_N} \mathbb{1}_R \ket{\Psi_N} = 1$, even in the thermodynamic limit. Thus, this construction has so far modeled a type II$_1$ algebra, such as that in de Sitter space. This can be modified to model a type II$_{\infty}$ algebra by appending a single Harmonic oscillator to the right side, which has Hilbert space $L^2(\mathbb{R})$. With this addition, there is no normalizable state in the Hilbert space that acts a trace. Nevertheless, the above identifications of $\mathcal{Z}'_{\mathcal{M}}$ and $\mathbb{1}_{obs}^{\mathcal{M}}$ go through, with the modification that $\rho_R \rightarrow \rho_R\otimes \mathbb{1}_{L^2(\mathbb{R})}$, which is not normalizable because it has an infinite trace.

The final step is generalizing this construction to a direct sum Hilbert space, labeled by $\mathcal{M}$. In this case, the number of qubits $N$ in each superselection sector is $\mathcal{M}$ dependent, i.e.~it can take different values in the asymptotic expansion in $N$. For simplicity, consider the case of just two sectors, though it is immediately generalizable.
The identity matrix on the global Hilbert space restricted to $R$ is 
\begin{align}
\begin{aligned}
 \left(\mathbb{1}_{R,\mathcal{M}_1}\otimes \mathbb{1}_{L,\mathcal{M}_1} \right)\oplus \left(\mathbb{1}_{R,\mathcal{M}_2}\otimes \mathbb{1}_{L,\mathcal{M}_2}\right)
 &= 2^{N(\mathcal{M}_1)}\tilde{\rho}_{R,\mathcal{M}_1}\oplus2^{N(\mathcal{M}_2)}\tilde{\rho}_{R,\mathcal{M}_2} 
\end{aligned}
\end{align}
where we have defined the density matrices such that $\Tr_R(\rho_R) = 1$. $\tilde{\rho}_{R,\mathcal{M}_1}$ and $\tilde{\rho}_{R,\mathcal{M}_2}$ are the analogs of $\mathbb{1}_{obs}^{\mathcal{M}_1}$ and $\mathbb{1}_{obs}^{\mathcal{M}_2}$. The prefactors are the analogs of $\mathcal{Z}'_{\mathcal{M}_1}$ and $\mathcal{Z}'_{\mathcal{M}_2}$.

\subsection{The path integral approach}
We now present a path integral argument along the lines of section \ref{sect:2.1}. One can consider the density matrix associated with a subregion if the subregion of interest is invariantly defined on a partial Cauchy surface. There are two general ways we know how to do this: (1) When the region is the causal diamond associated with an infinitely extended observer. In this case, the position of the observer at infinity invariantly define the region and is robust to perturbations of the geometry. This is the approach of \cite{witten2024background}. (2) When the partial Cauchy surface is bounded by an extremal surface, the subregion is unambigously defined because the extremal surface is robust to perturbations. This is the approach of \cite{Chen:2024rpx}. We will propose a no-boundary prescription that applies equally well in both cases.

When an observer sees only a portion of the spacetime, there is no way to prepare a pure quantum state. Therefore, we begin with physical boundary conditions for a density matrix where the bra and ket meet at a codimension-two surface, $\gamma$, which is the intersection of the boundaries of the observer's future and past light cones. In the case where $\gamma$ is an extremal surface, the observer's end-points need not be asymptotic and the causal diamond associated to the extremal surface need not be associated to any one observer worldline. With the causal diamond invariantly defined, using options (1) or (2), we then specify boundary conditions on these light cones
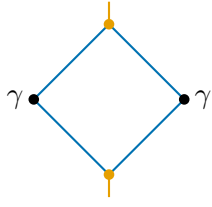
\begin{align}
     \begin{tikzpicture}
    \draw[thick,\bracolor] (0,0 ) -- (1,1)--(2,0) --(1,-1)--cycle;
     \fill[\obscolor] (1,1) circle (2pt); 
     \fill[\obscolor] (1,-1) circle (2pt); 
     \fill[] (0,0) circle (2pt);
      \fill[] (2,0) circle (2pt);
      \draw[thick, \obscolor] (1,1) -- (1,1.3);
      \draw[thick, \obscolor] (1,-1) -- (1,-1.3);
      \node[] at (-.25 , 0) {$\gamma$};
       \node[] at (2.25 , 0) {$\gamma$};
\end{tikzpicture}
\end{align}
The purpose of using null surfaces instead of spatial surfaces for the boundary conditions is so that the fact that the observer only sees a portion of the universe is explicitly part of the definition of the path integral.\footnote{We thank Edward Witten for suggesting null surfaces as the appropriate partial Cauchy slices for boundary conditions.
} 
The surface $\gamma$ may be connected or not depending on the background. For simplicity, we will restrict ourselves to cases where the topology of these null boundaries is the same as the null boundaries of the static patch in de Sitter. More complicated topologies require more care, though we expect that none of our conclusions are sensitive. Matrix elements of the observer's no-boundary state are computed by path integrating over all geometries and field configuration $\phi_\text{bulk}$ consistent with boundary conditions on these null surfaces, as in \eqref{eq:nb_density_filled}\footnote{The notation $\ket{\phi'}$ and $\ket{\phi''}$ is formal because we do not expect to have a genuine Hilbert space for a subregion.} 
 \begin{align}
 \bra{\phi''}\rho_\text{NB}\ket{\phi'}=\int \mathcal{D}\phi_\text{bulk}\quad 
 \raisebox{-0.47\height}{\begin{tikzpicture}
    \node at (-1.3,0) {$\phi_\text{bulk}$}; 
    \draw[thick,\bracolor] (0,0 ) -- (1,1)--(2,0) --(1,-1)--cycle;
     \fill[\obscolor] (1,1) circle (2pt); 
     \fill[\obscolor] (1,-1) circle (2pt); 
     \fill[] (0,0) circle (2pt);
      \fill[] (2,0) circle (2pt);
    \draw[thick] (1,0) circle (1.5);
    \draw[thick, \obscolor] (1,1) to[out = 90,in= 90](-.25,0) to[out=-90,in=-90] (1,-1); 
    \node at (1.75,0.75) {$\phi'$};
    \node at (1.75,-.75) {$\phi''$};
\end{tikzpicture}}\label{4.17rho}
\end{align}
This causal diamond is well-defined semiclassically if we consider a subspace of states that all have the same expectation value of induced metric and extrinsic curvature on this diamond at $O(1)$, but are generally different at $O(\sqrt{\hbar G})$ and have small fluctuations consistent with the uncertainty relations $\braket{\Delta K \Delta g}\lesssim \hbar G$. In this subspace, the expectation values provide initial data that fully specifies the geometry at leading order within the causal diamond, without specifying the geometry outside of the diamond. In practice, one can complete the partial Cauchy surface to a full Cauchy surface, and place ``trace-out'' boundary conditions in the unobservable region \cite{Ivo:2024ill}.

Matrix elements of $\rho_\text{NB}^2$ are computed along the lines of equation \eqref{eq:nb_density_squared} as
\begin{align}
\bra{\phi''}\rho_\text{NB}^2\ket{\phi'} = \underset{\substack{\text{boundary conditions}\\\text{modulo diffeomorphisms}}}{\int\mathcal{D}\psi}\,\,\raisebox{-0.46\height}{\begin{tikzpicture}
    \draw[thick,\bracolor] (0,0 ) -- (1,1)--(2,0);
    \draw[thick,\bradiffcolor] (0,0 ) -- (1,-1)--(2,0);
     \fill[\obscolor] (1,1) circle (2pt); 
     \fill[\obscolor] (1,-1) circle (2pt); 
     \fill[] (0,0) circle (2pt);
      \fill[] (2,0) circle (2pt);
    \draw[thick] (1,0) circle (1.5);
    \draw[thick, \obscolor] (1,1) to[out = 90,in= 90](-.25,0) to[out=-90,in=-90] (1,-1); 
     \draw[thick,\bradiffcolor] (0+3.5,0 ) -- (1+3.5,1)--(2+3.5,0);
    \draw[thick,\bracolor] (0+3.5,0 ) -- (1+3.5,-1)--(2+3.5,0);
     \fill[\obscolor] (1+3.5,1) circle (2pt); 
     \fill[\obscolor] (1+3.5,-1) circle (2pt); 
     \fill[] (0+3.5,0) circle (2pt);
      \fill[] (2+3.5,0) circle (2pt);
    \draw[thick] (1+3.5,0) circle (1.5);
    \draw[thick, \obscolor] (1+3.5,1) to[out = 90,in= 90](-.25+3.5,0) to[out=-90,in=-90] (1+3.5,-1); 
    \node at (1.75,0.75) {$\phi'$};
    \node at (1.8,-.75) {$\psi$};
    \node at (1.75+3.5,0.75) {$\psi$};
    \node at (1.75+3.5,-.75) {$\phi''$};
\end{tikzpicture}}\,\,\,=\bra{\phi''}\rho_\text{NB}\ket{\phi'}\,.\label{4.19rhosq}
\end{align}
The matrix multiplication is not a simple geometric gluing of the green surfaces. Rather, similar to the inner product for full Cauchy slices, it involves integrating over geometries with the past boundary conditions given by the green surface in the left diagram and the future boundary conditions given by the green surface in the right figure. 

A crucial point is that matrix multiplication does not result in a conical excess at $\gamma$. This is distinct from the replica trick in AdS/CFT where a conical excess arises at the asymptotic boundary where the two copies are glued. In a closed universe, there is nothing to enforce the conical excess like the non-gravitating boundary where the CFT resides, and gravity smooths out the singularity. The integral over bulk geometries, which was left implicit in the picture, selects the smooth single-cover configuration. An analogous comment was made in \cite{Ivo:2024ill} in the context of the no-boundary density matrix for slow-roll inflation. We conclude from \eqref{4.19rhosq} that
\begin{equation}
 \rho_\text{NB}^2=\rho_\text{NB}\,.
\end{equation}
Therefore, the observer's no-boundary density matrix is the identity matrix for this subregion.

As in section \ref{sect:2.2}, the density matrix prepared by the path integral is not normalized. The partition function 
$\mathcal{Z}'_{\mathcal{M}}$ 
is the norm of the density matrix \eqref{4.17rho}.\footnote{For example, in de Sitter space, the matrix multiplication involves a purely Lorentzian evolution of both the bra and ket to the spatial $t = 0$ slice, before gluing. The Lorentzian parts of the path integral cancel. The partition function is then dominated by the Euclidean action of the sphere.} In the sense of an asymptotic expansion in $G$, this determines the additive constant in the von Neumann entropy of normalized states $\rho$ in $\mathcal{H}_{\mathcal{M}}$
\begin{align}
 S(\rho) = \log \mathcal{Z}'_{\mathcal{M}} +O(1).
\end{align}
In the case of dS, the sphere path integral equals to leading order the area of the cosmological horizon. In more general spacetimes where the region of interest is not bounded by an extremal surface, we do not expect that the path integral is equal to the area. We comment more on this in a specific example in the discussion section \ref{sect:5.conclusion}.\footnote{This is somewhat different to the conclusions of \cite{Jensen:2023yxy} where entropy changes were related to area changes of the boundary of the subregion. We attribute this difference (in the case of non-extremal surfaces) to a difference in how we are defining our subregions. We define the subregion as the accessible region for an observer, whereas \cite{Jensen:2023yxy} appear to be making a ``gauge choice'' not related to any particular observer's causal diamond, the same as in \cite{Chandrasekaran:2019ewn}. With their choice, the location of the bifurcation surface is unchanged under perturbations of spacetime. In the case where the surface is extremal, the two definitions agree.} Interestingly, the result for the partition function in \cite{Ivo:2024ill} is somewhere in between. It is not the (very large) area of the boundary of the region of interest but rather the area of a surface to the past of the region, namely, the horizon at the end of inflation.

We can compare our analysis to that of \cite{Chen:2024rpx} for regions bounded by an extremal surface. In \cite{Chen:2024rpx}, a gravitational mode related to the area operator that generates a local boost about the extremal surface was quantized, leading to a type II$_{\infty}$ algebra. One expects this algebra to not have a center, so the trace is unique up to rescalings. We have shown that the no-boundary state is also tracial for such an algebra. Thus, the tracial state from \cite{Chen:2024rpx} ought to be the $O(1)$ contribution to our no-boundary state for these subregions. Indeed, in the specific example of slow-roll inflation, \cite{Chen:2024rpx} explicitly invoked the semiclassical limit of the no-boundary state to land on a trace. It would be extremely interesting if there was an analogous gravitational mode for non-extremal surfaces to more explicitly analyze entropy in these spacetimes. We expect that if such a mode exists, it will include corrections to the area operator. That would provide a complementary explanation for why the entropy is related to action more generically. Such an analysis may be tractable in low-dimensional models, extending \cite{Kaplan:2022orm,Dong:2025orj}.

The global no-boundary state with an observer has no contribution from a sphere topology. However, when reducing the no-boundary state to a state for a subregion by tracing out ``degrees of freedom'' in the other patch, the path integral does generally include a sphere topology. This results in a confusing formal equality

\begin{align}
    \begin{tikzpicture}
    \node[] at (-3.2,-.5) {$\Tr_{\text{other patch}}$};
    \draw[thick, \bracolor] (0,0) ellipse (1 and .25);
    \draw[thick, \bracolor] (0,-1) ellipse (1 and .25);
    \draw[thick ] (-1,0) to[out = 90, in = 90] (-1.5,-.5) to[out = -90, in = -90](-1,-1) ;
    \draw[thick ] (1,0) to[out = 90, in = 90] (-2,-.5) to[out = -90, in = -90](1,-1) ;
    \draw[thick ,\obscolor] (0,.25) to[out = 90, in = 90] (-1.75,-.5) to[out = -90, in = -90](0,-1.25) ;
    \node[] at (1.75,-.5) {$\overset{?}{=}$} ;
    \draw[thick, black ] (4,-.5) ellipse (1.5 and .75);
    \draw[thick] (3.75,-.5)to[out = -30, in = -150](4.25,-.5) ; 
    \draw[thick] (3.85,-.5)to[out = 30, in = 150](4.15,-.5) ; 
    \draw[\obscolor, thick] (4.75,-.45) to[out= 90, in = 90] (3.25,-.5) to[out = -90, in = -90] (4.75,-.55);
    \draw[\bracolor, thick] (4.35,-.5) to[out = 20,in= 160] (5.15,-.5);
    \draw[\bracolor, thick] (4.35,-.5) to[out = -20,in= -160] (5.15,-.5);
    \node[] at (6,-.5) {$+$} ;
    \draw[black, thick] (7.75,-.5) circle (1.25);
    \node[] at (10,-.5) {$+\quad \dots$} ;
    \draw[thick, \obscolor] (7.95,-.4)to[out = 90, in = 90](6.75,-.5) to[out = -90, in = -90](7.95,-.6);
        \draw[thick, \bracolor] (7,-.5) to[out = 10,in= 170] (8.9,-.5);
    \draw[thick, \bracolor] (7,-.5) to[out = -10,in= -170] (8.9,-.5);

        \fill[\obscolor] (0,.25) circle (2pt); 
     \fill[\obscolor] (0,-1.25) circle (2pt); 

            \fill[\obscolor] (4.75,-.4) circle (2pt); 
     \fill[\obscolor] (4.75,-.6) circle (2pt); 
        \fill[\obscolor] (7.95,-.4) circle (2pt); 
     \fill[\obscolor] (7.95,-.6) circle (2pt); 

\end{tikzpicture}
\end{align}
This seems clear from the rules of the gravitational path integral, but is non-perturbative. A detailed understanding of this would thus require understanding the decomposition of the Hilbert space on the global slice into Hilbert spaces for subregions at the non-perturbative level. Assuming the equality makes sense, it would be interesting to understand how the sphere emerges, but this is beyond our current scope.
Perhaps the simplest theory to investigate this is JT gravity with $\Lambda > 0$. There, the wormhole amplitude without an observer computed in \cite{Cotler:2019dcj,Cotler:2024xzz} is an identity operator \cite{Held:2024rmg}, but a semiclassical observer has restricted causal access, and so the observer's density matrix had not been fully understood thus far. A similar observation was also made in \cite{Fumagalli:2024msi}.

\section{No KMS theorem}
\label{sect:nogo}
We would like to check in controllable examples that the density matrix for an observer with causal horizons in section \ref{sect:3obshorizon} defines a trace. A case in which it is understood how to find the trace is when the matter state satisfies the Kubo–Martin–Schwinger (KMS) condition with respect to time translations on an observer's worldline. In this case, the matter can be dressed to the observer and an explicit formular for the trace can be found \cite{witten2024background}.
In de Sitter space, there exists a state that is KMS in the full static patch, the so-called Bunch-Davies state \cite{Chernikov:1968zm,schomblond1976conditions,Bunch:1978yq,Mottola:1984ar,allen1985vacuum}, which was critical the analysis of CLPW. In this section, we show that this construction cannot work for any other spacetime that is asymptotically de Sitter in the far past or future. In particular, we show that the existence of a KMS state, even when we only require it to be KMS on the observer's worldline, is unique to de Sitter.\footnote{For some interesting results on the sparsity of KMS states, see \cite{kay1991theorems, Keyl:1993ye, Borchers:2000pv,Pinamonti:2018ltu,Sorce:2024zme}.} We emphasize that this does not imply that the no-boundary state is not the trace or that a trace cannot be defined. It simply means that the explicit construction of the trace requires more work than the construction of \cite{witten2024background} and that the resulting entropy cannot describe an equilibrium thermodynamic entropy. We discuss promising directions on this front in section \ref{sect:5.conclusion}.

Before presenting the arguments, we explain what is meant by a KMS state on the worldline. Given a quantum state and a worldline geodesic parametrized by a time $t$, we may consider correlation functions of the form
\begin{equation}\label{eq:some_corr}
 f(t_1,t_2) = \langle \phi_1(t_1) \phi_2(t_2) \rangle\,,
\end{equation}
where $\phi_{1,2}$ are any local operators in the theory.
The KMS property of the state requires that $f(t_1+z, t_2)$ is analytic for in the interior of the strip $-\beta < \Im(z) < 0$ and
\begin{equation}\label{eq:kms_1}
 \langle \phi_1(t_1) \phi_2(t_2) \rangle = \langle \phi_2(t_2) \phi_1(t_1+\i\beta)\rangle\,.
\end{equation}
This equation (or the requirement for analyticity) is not required to hold for operators not on the worldline.

We note that a (faithful semi-finite normal) state always satisfies the KMS condition with respect to abstract modular time evolution. In this section, we are concerned with the more restrictive condition that this modular evolution coincides with a physical time evolution.

\subsection{General argument}\label{sect:5.1kms}
Let us now assume that the perturbative algebra of observables on a worldline in an asymptotically de Sitter background admits a KMS state with respect to proper time translations on the worldline. In particular, we consider FLRW spacetimes of the form
\begin{align}
\label{eq:frwmetric}
 \d s^2 = -\d t^2 + a(t)^2\d\Omega_{3}\,,\quad a(t) \sim \frac{1}{H}e^{H |t|} \text{ for } \abs{t}\to \infty\,,
\end{align}
where $d\Omega_3$ is the metric on the unit three-sphere.
We will show that the state on the observer's worldline in such a geometry can only satisfy KMS if the geometry is exactly de Sitter space. To prove this, we first show that if correlation functions on the worldline satisfy KMS, they are indistinguishable from the correlation functions of the Bunch-Davies state. We then show that the Bunch-Davies state is inconsistent with the operator product expansion (OPE) for any background except for de Sitter.

Consider the two-point function on the worldline \eqref{eq:some_corr}. 
Because there are no light-cone singularities on the worldline, the correlation function is analytic for a given ordering $t_1 \ne t_2$. If it satisfies KMS \eqref{eq:kms_1}, it is periodic under a combined shift $t_i \to t_i + \i \beta$ and is analytic on the strip $0\le \Im (t_i) \le \beta$. Consider now 
\begin{equation}
 f(T) = \langle \phi_1(t_1+T) \phi_2(t_2+T) \rangle\,.
\end{equation}
This agrees with \eqref{eq:some_corr} at $T=0$.
By the above assumptions, this function is periodic $f(T)=f(T+\i \beta)$ and analytic on the strip $0\le \Im (T) \le \beta$. Because the background is asymptotically de Sitter, the correlation function asymptotes to the Bunch-Davies correlation function at large $|T|$, by the ``cosmic no-hair theorem'' \cite{wald1983asymptotic,Hollands:2010pr,Carroll:2017kjo,Marolf:2010zp,Marolf:2010nz}.
This is a finite limit.
By Liouville's theorem, $f(T)$ is thus a constant as a function of $T$. This shows that the correlation function agrees with the BD correlation function on the worldline for all times.

Every state must respect the OPE of the theory. The OPE is a property of the algebra and depends on the underlying background geometry \cite{Hollands:2014eia}. It places constraints on possible correlation functions. While the Bunch-Davies state satisfies these constraints for the de Sitter background, it will not satisfy them, even only on the worldline, for any other background.
To illustrate this point, consider a massless minimally coupled Klein-Gordon scalar field on an FLRW background.\footnote{An analogous argument holds when adding mass or considering the free Maxwell field or linearized graviton, and for more general geometries.} Free fields in curved spacetime satisfy the canonical commutation relations \cite{Hollands:2014eia}
\begin{equation}
 [\phi(x) ,\phi(y)] = \i E(x,y)\ \mathbb{1}\,.
\end{equation}
Here, $E(x,y)$ is the advanced minus retarded Green's function for the Klein-Gordon equation. Therefore every state in the theory must have the property that correlation functions on the worldline satisfy
\begin{align} \label{eq:comm_cond}
 \braket{[\phi(t_1) ,\phi(t_2)] }= \i E(t_1,t_2)\,.
\end{align}
For $t_1\to t_2$, the Green's function takes a universal form \cite{dewitt1960radiation,kay1991theorems}
\begin{align}\label{eq:exp_E}
 E(t_1,t_2) \to \frac{R(t_1)}{48\pi}+\dots,\quad t_1 < t_2.
\end{align}
However, correlation functions in the Bunch-Davies state are (by definition) identical to dS correlation functions and satisfy \cite{allen1985vacuum}
\begin{align}\label{eq:comm_ds}
 \braket{[\phi(t_1) ,\phi(t_2)] } = \i \frac{H^2}{4\pi}, \quad t_1 < t_2.
\end{align}
This is inconsistent with \eqref{eq:exp_E} for spacetimes without constant scalar curvature. Thus, the Bunch-Davies state for FLRW spacetimes is not in the Hilbert space. Higher orders in the OPE expansion \eqref{eq:exp_E} depend on all the curvature components \cite{Hollands:2023txn,Hollands:2006ag}, so even for inhomogeneous backgrounds, the BD constraint $E=H^2/4\pi$ fixes the background around the worldline to that of de Sitter.
Under suitable conditions on the metric, this will constrain the geometry for the entire region accessible to the observer. We expect that similar arguments using the OPE as a constraint should hold for other local quantum field theories.
In summary, we conclude that the only asymptotically de Sitter spacetime that admits a KMS state on a timelike worldline is de Sitter space itself. 

This statement is somewhat related to the cosmic no-hair theorem. Indeed, in order to define KMS states in QFT, the geometry should possess a timelike isometry which acts as time translations on the worldline.\footnote{For conformally invariant QFTs, a timelike isometry may be replaced by a timelike conformal isometry.} This requires the background geometry in the causal region of the observer to be stationary. Since the geometry is also asymptotically de Sitter, the cosmic no-hair theorem fixes the geometry in the causal region to de Sitter. 
\subsection{Constructive argument}\label{sect:5.2constructive}
In this section, we flip the logic around 
by attempting to construct FLRW solutions with conformal matter correlation functions on a worldline that satisfy KMS. We will find that only de Sitter space satisfies this requirement. 

We again consider the FLRW metric \eqref{eq:frwmetric}, now 
coupled to a conformal field theory with a primary operator $O_\Delta$ of dimension $\Delta$. Using the conformal time coordinate $\d t = a \ d\eta$, the metric is related to $R\times S^3$ as
\begin{equation}
 \d s^2 = a(t)^2(-\d \eta^2 + \d \Omega_3)\,.
\end{equation}
Time translation is therefore a conformal Killing direction in this background. Consider the state that annihilates the time translation operator.
The conformal two-point function on this state is therefore
\begin{equation}\label{conf2p_flrw}
 \langle O_\Delta(t_1,0) O_\Delta(t_2, \alpha)\rangle = \frac{1}{a^\Delta(t_1) a^\Delta(t_2)} G_\Delta(\eta(t_1)-\eta(t_2),\alpha)\,,
\end{equation}
with $\alpha$ the angular difference in $S^3$ and $G_\Delta(\eta,\alpha)$ the vacuum conformal two-point function on $R \times S^3$
\begin{equation}
 G_\Delta(\eta_1,\eta_2,\alpha)=\left(2\cos(\eta_1-\eta_2)-2\cos \alpha\right)^{-\Delta}\,.
\end{equation}
A necessary condition for \eqref{conf2p_flrw} to satisfy KMS is that it is analytic in the complex $t_1$ strip away from the usual short-distance singularities. $G_\Delta(\eta,\alpha)$ has this property, so for analyticity we need $\eta(t)$ and $a(t)$ to be analytic.
Moreover, the two-point function has a branch cut whenever $a(t)=0$ with $|\eta(t)|<\infty$. 

The most general analytic function $a(t)$ that is also periodic $a(t+i\beta) = a(t)$ (as demanded by KMS) is a sum of $\sinh(n \, H t)$'s and $\cosh(n \, H t)$'s with integer $n$ and $H=2\pi/\beta$. The asymptotic de Sitter condition imposes that $a(t) \simeq e^{H|t|}$ for large $|t|$, restricting $n$ to $0$ or $1$. Choosing $t=0$ as a moment of time-reflection symmetry, one is left with the functions $a(t) = a_0 + a_1 \cosh(H t)$. Wick-rotating to $\tau = \i t$, there exists an Euclidean time coordinate $\tau=\tau_0$ with $a(\pm\tau_0) = 0$. To avoid a conical singularity at this point, one should impose $\dot a(\pm\tau_0) = \pm 1$. The most general analytic solution is thus found to be 
\begin{equation}\label{eq:t0_sol}
 a(t) = \frac{\cosh(H t)-\cos(H\tau_0)}{H \sin(H\tau_0)}.
\end{equation}
This FLRW geometry can actually be realized by considering a positive cosmological constant $\Lambda = 3H^2$ and a homogeneous density of domain walls\footnote{These generically appear in models of the early universe when there is spontaneous symmetry breaking via the Kibble-Zurek mechanism \cite{kibble1976topology}. A physically motivated case arises in the invisible axion extension of the Pecci-Quinn model \cite{shifman1980can,dine1981simple,Zhitnitsky:1980tq}. At energies below the QCD scale, domain walls form in a network, connecting cosmic strings \cite{sikivie1982axions}. These are topologically stable and are the source of the domain wall problem \cite{zel1974cosmological}.}, which have energy density redshifting as $\omega/a$. An exact solution to Einstein's equations is then indeed of the form \eqref{eq:t0_sol}
\begin{figure}
 \centering
 \begin{tikzpicture}

        \draw[thick] (-3,0) ellipse (1 and .25);
    \draw[thick] (0-3,-2) ellipse (1 and .25);
    \draw[] (0-3,-1) ellipse (.5 and .125);
    \draw[thick] (-1-3,0) to[out = -70, in = 90] (-.5-3,-1) to[out = -90, in = 70] (-1-3,-2) ;
    \draw[thick] (1-3,0) to[out = -110, in = 90] (.5-3,-1) to[out = -90, in = 110] (1-3,-2) ;

    \draw[thick] (0,0) ellipse (1 and .25);
    \draw[thick] (0,-2) ellipse (1 and .25);
    \draw[] (0,-1) ellipse (.1 and .025);
    \draw[thick] (-1,0) to[out = -70, in = 90] (-.1,-1) to[out = -90, in = 70] (-1,-2) ;
    \draw[thick] (1,0) to[out = -110, in = 90] (.1,-1) to[out = -90, in = 110] (1,-2) ;

    \draw[thick] (3,0) ellipse (1 and .25);
    \draw[thick] (3,-2) ellipse (1 and .25);
    \draw[] (3,-1) ellipse (.01 and .005);
    \draw[thick] (2,0) to[out = -70, in = 90] (-.01+3,-1) to[out = -90, in = 70] (2,-2) ;
    \draw[thick] (4,0) to[out = -110, in = 90] (3.01,-1) to[out = -90, in = 110] (4,-2) ;
\end{tikzpicture}
 \caption{Cartoon of Lorentzian spacetimes \eqref{5.24} when $\omega = 0$ (left), $0< \omega < \infty$ (middle), and $\omega \rightarrow \infty$ (right), each with de Sitter asymptotics. The throat shrinks as $\omega$ increases, pinching off in the $\omega \rightarrow \infty$ limit. }
 \label{fig:pinchoff_tikz}
\end{figure}
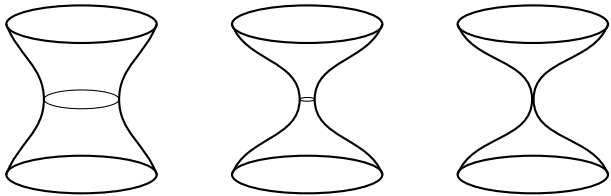
\begin{equation}
 a(t)=\frac{1}{2H^2}\left(\sqrt{4H^2+\omega^2}\cosh(H t)-\omega\right)\,.\label{5.24}
\end{equation}
These solutions are pictured in figure \ref{fig:pinchoff_tikz}. For $\omega=0$, the solution reduces to pure de Sitter. The length of the throat goes to zero in the extreme $\omega\to\infty$ limit. 

To appreciate that the analyticity of $a(t)$ goes some way towards having a KMS state, 
we have checked explicitly that on the worldline ($\alpha=0$) 
in the geometry \eqref{5.24}, correlation functions of scalar
conformal primaries \eqref{conf2p_flrw} are equal to those of the Bunch-Davies state for de Sitter with $\beta=2\pi/H$. 
Unfortunately (but in line with the discussion of section \ref{sect:5.1kms})
however, even this spacetime with analytic $a(t)$ does not result in a KMS state on the worldline for all operators. To see this, notice that the Ricci scalar of \eqref{eq:t0_sol} is given by
\begin{align}\label{eq:ricci_calc}
 R =6 H^2\frac{2 \cosh (H t)+\cos (H\tau_0)}{\cosh (H t)-\cos (H\tau_0)}\,.
\end{align}
The curvature thus diverges at the poles of the Euclidean geometry $t=\pm \i \tau_0$ unless $\tau_0 = \pi/2H$, which is the de Sitter ($\omega=0$). As a result, some correlation functions are non-analytic. For instance, consider a conformally-coupled scalar $\phi$. While the associated primary two-point function is equal to the Bunch-Davies correlation function, the Laplacian descendant is divergent at the poles\footnote{By the Klein-Gordon equation, the prefactor is indeed proportional to the Ricci scalar.}
\begin{equation}
 \langle \nabla^2 \phi(t_1) \phi(t_2)\rangle = H^2 \frac{2\cosh(H t_1)+\cos(H \tau_0)}{\cosh(H t_1)-\cos(H \tau_0)} \, \langle \phi(t_1) \phi(t_2)\rangle.
\end{equation}
The only FLRW geometry that avoids this issue and thus has the potential for worldline KMS is de Sitter itself with $a(t)=\cosh(Ht)/H$, as we argued also in section \ref{sect:5.1kms}.

\section{Concluding remarks}\label{sect:5.conclusion}
We have argued using the gravitational path integral that the no-boundary density matrix of the universe with an observer is the identity matrix on the physical Hilbert space and thus provides a trace for the observer's algebra. This realizes a proposal of \cite{witten2024background} that the observer's no-boundary state can be used to define von Neumann entropy in a background-independent manner so that there is no arbitrary constant for each spacetime. To provide a concrete example, we showed that Euclidean wormhole amplitudes in JT gravity (i.e.\,the no-boundary path integral) compute matrix elements of an identity operator on the physical Hilbert space.

\begin{figure}
 \centering
 \def\dy{.1}
\usetikzlibrary{decorations.pathmorphing}
\begin{tikzpicture}

    \draw[dashed] (0,0) to (4,0);
    \draw[dashed] (0,-2) to (4,-2);
    \draw[dashed] (2,0) -- (2,-2);
    \draw[->, decorate, decoration={zigzag, segment length=6, amplitude=2}, color=\bracolor] (0,-1) -- (1,-1);
    \draw[->, decorate, decoration={zigzag, segment length=6, amplitude=2}, color=\bracolor] (3,-1) -- (4,-1);
    \draw[->,very thick,\obscolor] (2+\dy,-2+\dy) -- (3.5,-2+\dy); 
    \draw[->,very thick,\obscolor] (2+\dy,-2+\dy) -- (2+\dy,0+\dy) -- (3.5,0+\dy); 
    
    \fill (3.5,0) circle (.05) node[above]{$\phi(T+\i\beta)$};
    \fill (3.5,-2) circle (.05) node[below]{$\phi(T)$};
    \fill[\obscolor] (2,-2) circle (.05) node[below,black]{$0$};
\end{tikzpicture}
 \caption{The complex $t$ plain for the axion wormhole \eqref{6.3}. The dashed lines represent where the correlation functions are analytic. The horizontal line is the Lorentzian contour and the vertical line is the Euclidean contour. The existence of a smooth Euclidean background gives a path (in orange) which allows one to define $\phi(T+\i \beta)$ unambiguously. The correlation functions of these fields satisfy ``weak KMS'' \eqref{eq:kms_1}.}
 \label{fig:weak_kms}
\end{figure}
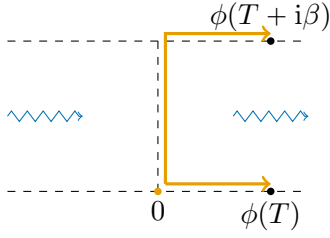

Following the claims of section \ref{sect:3obshorizon} where the observer has causal horizons, one would like to explicitly solve for the no-boundary state on a specific background.
The no-go results of section \ref{sect:nogo} suggest that 
a generic no-boundary saddle does not prepare a KMS state on the worldline which is an obstruction to repeating the analysis of \cite{Chandrasekaran:2022cip,witten2024background}. 
It can, however, prepare a state with a ``weak KMS'' relation. By this we mean that from the no-boundary prescription, the observer worldline has the topology of a circle in the complex geometry. On an open set around the worldline, quantum field theory correlation functions remain analytic (see figure \ref{fig:weak_kms}). This allows one to uniquely define an analytic continuation of Lorentzian operators of the type $\phi(t+\i \beta)$ that can be shown to satisfy a KMS-like relation \eqref{eq:kms_1}. We stress that, unlike in the standard KMS relation, we do not have analyticity on the entire strip $0\le \Im (t) \le \beta$. It is plausible that this ``weak KMS'' \eqref{eq:kms_1} 
suffices to explicitly find the trace for these backgrounds as well. We are obligated to mention the logical possibility that there is no perturbative trace on a general background, though we are optimistic that this is not the case. In this case, the observer's algebra would be type III. This is not in contradiction with any of our claims of section \ref{sect:3obshorizon}, as mentioned in footnote \ref{footnote}. 

We conclude this paper with a digestif, by introducing a particular class of spacetimes where one can test our proposal. We intend to report more details regarding this elsewhere.
We consider solutions of the Friedmann equations with a positive cosmological constant supported by axion flux \cite{Aguilar-Gutierrez:2023ril}. Recall the 3D FLRW metric\footnote{It is straightforward to generalize to higher dimensions, though the formulas become slightly less elegant.}
\begin{equation}
\label{FLRW_coord}
 \d s^2=-\d t^2+ a(t)^2\d \Omega_2\,,\quad \d \Omega_2=\d\theta^2+\sin^2(\theta)\d \phi^2\,.
\end{equation}
The Friedmann equation with axion flux density $Q \equiv q H^{-1}({16\pi G})^{-1/2}$
becomes
\begin{align}
 \frac{1}{a^2}\bigg(\frac{\d a}{\d t}\bigg)^2 = -\frac{1}{a^2} +H^2 + \frac{q^2}{4 H^2}\frac{1}{a^4}\,\label{5.3friedmann}
\end{align}
and solve to
\begin{equation}
 a(t)=\left(\frac{1+\sqrt{1-q^2}\cosh(2 Ht)}{2H^2}\right)^{1/2}\,.\label{6.3}
\end{equation}
This solution has a minimal size at $t=0$ and asymptotes to de Sitter space for $\abs{t}\to\infty$ if $q<1$. If $q = 1$, this is an Einstein static universe $a=\tfrac{1}{\sqrt{2}H}$. To check whether a comoving observer in this spacetime has causal horizons, we compute the conformal time that passes between the $t = 0$ and $t=\infty$. For an observer at $\theta=0$, one finds that there is a causal horizon at\footnote{The metric can be brought in conformal form by the coordinate transformation
\begin{equation}
 \eta(t)=\int_{0}^t \frac{\d t}{a(t)}.
\end{equation}
Half of the height of the causal diamond in the Penrose diagram is obtained in the limit $t\to+\infty$ of the observer worldline. 
}
\begin{equation}
 \theta_\text{h}(q)=\sqrt{\frac{2}{1+\sqrt{1-q^2}}} \ 
 K\bigg(\frac{1-\sqrt{1-q^2}}{1+\sqrt{1-q^2}}\bigg),
\end{equation}
where $K$ is the complete elliptic integral of the first kind.
In empty de Sitter space ($q=0$), a static observer sees exactly half of the space $\theta_\text{h}=\tfrac{\pi}{2}$. For $q>0$, an observer sees more than half of the spacetime, consistent with the general fact that positive energy brings the antipodal observers in dS in causal contact with one another \cite{Gao:2000ga}. 
Interestingly, there is a critical value $q_\text{c}$ such that for $q>q_\text{c}$ the observer has complete causal access. Therefore, the observer's algebra is expected to transition to a type I von Neumann algebra. 
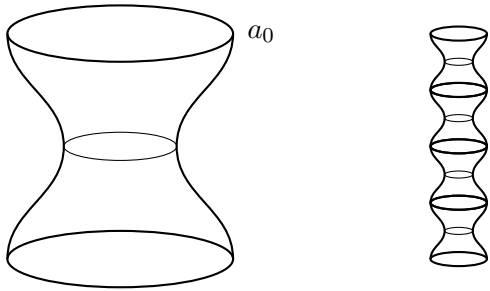
\begin{figure}
 \centering
 \begin{tikzpicture}[scale = 1.5]

    \draw[thick] (-3,0) ellipse (1 and .25);
    \draw[thick] (0-3,-2) ellipse (1 and .25);
    \draw[] (0-3,-1) ellipse (.5 and .125);
    \draw[thick] (-1-3,0) to[out = -90, in = 90] (-.5-3,-1) to[out = -90, in = 90] (-1-3,-2) ;
    \draw[thick] (1-3,0) to[out = -90, in = 90] (.5-3,-1) to[out = -90, in = 90] (1-3,-2) ;
    \node[] at (-1.75,0) {$a_0$};

    \draw[thick] (0,0) ellipse (1/4 and .25/4);
    \draw[thick] (0,-2/4) ellipse (1/4 and .25/4);
    \draw[] (0,-1/4) ellipse (.5/4 and .125/4);
    \draw[thick] (-1/4,0) to[out = -90, in = 90] (-.5/4,-1/4) to[out = -90, in = 90] (-1/4,-2/4) ;
    \draw[thick] (1/4,0/4) to[out = -90, in = 90] (.5/4,-1/4) to[out = -90, in = 90] (1/4,-2/4) ;

    \draw[thick] (0,0-1/2) ellipse (1/4 and .25/4);
    \draw[thick] (0,-2/4-1/2) ellipse (1/4 and .25/4);
    \draw[] (0,-1/4-1/2) ellipse (.5/4 and .125/4);
    \draw[thick] (-1/4,0-1/2) to[out = -90, in = 90] (-.5/4,-1/4-1/2) to[out = -90, in = 90] (-1/4,-2/4-1/2) ;
    \draw[thick] (1/4,0/4-1/2) to[out = -90, in = 90] (.5/4,-1/4-1/2) to[out = -90, in = 90] (1/4,-2/4-1/2) ;

        \draw[thick] (0,0-1) ellipse (1/4 and .25/4);
    \draw[thick] (0,-2/4-1) ellipse (1/4 and .25/4);
    \draw[] (0,-1/4-1) ellipse (.5/4 and .125/4);
    \draw[thick] (-1/4,0-1) to[out = -90, in = 90] (-.5/4,-1/4-1) to[out = -90, in = 90] (-1/4,-2/4-1) ;
    \draw[thick] (1/4,0/4-1) to[out = -90, in = 90] (.5/4,-1/4-1) to[out = -90, in = 90] (1/4,-2/4-1) ;

    \draw[thick] (0,0-3/2) ellipse (1/4 and .25/4);
    \draw[thick] (0,-2/4-3/2) ellipse (1/4 and .25/4);
    \draw[] (0,-1/4-3/2) ellipse (.5/4 and .125/4);
    \draw[thick] (-1/4,0-3/2) to[out = -90, in = 90] (-.5/4,-1/4-3/2) to[out = -90, in = 90] (-1/4,-2/4-3/2) ;
    \draw[thick] (1/4,0/4-3/2) to[out = -90, in = 90] (.5/4,-1/4-3/2) to[out = -90, in = 90] (1/4,-2/4-3/2) ;
\end{tikzpicture}
 \caption{Left: The Euclidean FLRW wormhole geometry described in \eqref{5.8a}. Right: The necklace geometry.}
 \label{fig:kettlebell}
\end{figure}

The Lorentzian geometry can be continued ($\tau = \i t$) to a smooth no-boundary Euclidean wormhole solution (see figure \ref{fig:kettlebell})
\begin{equation}
 a(\tau)=\left(\frac{1+\sqrt{1-q^2}\cos(2H\tau)}{2H^2}\right)^{1/2}
 \,,\quad 0<H \tau<\pi .
 \label{5.8a}
\end{equation}
When $q\rightarrow 1$, we can see explicitly that the algebra of dressed observerables has a trace corresponding to the no-boundary state. This is because the path integral prepares a thermal state of quantum fields including gravitons with respect to global time translations. Using the KMS property of the quantum fields, the tracial property of the expectation value of dressed operators can be verified \cite{witten2024background}. In this case, one finds that the trace is the standard Hilbert space trace because the algebra is type I.

We have previously argued that the von Neumann entropy appearing in every state for the algebra on this background includes a contribution of minus the action of the no-boundary background \eqref{actionentropy4.11}. The Euclidean on-shell action is straightforward to compute \cite{Aguilar-Gutierrez:2023ril}
\begin{equation}
 I=-\frac{\pi }{{2 G}H}\left(1- q \right).
\end{equation}
This vanishes for $q=1$ which is the expected behavior for a type I algebra 
because type I algebras admit pure states with zero entropy.
The horizon area for this geometry is distinct from (minus) the action for $0<q<1$
\begin{align}
 A_h = 2\pi H^{-1}(1+\sqrt{1-q^2})\sin(\theta_\text{h})^2\,.
\end{align}
For $0<q<1$, one is tempted to interpret $-I$ as the leading contribution of the entropy, following section \ref{sect:3obshorizon}. This is also the answer one would arrive at from the gravitational replica trick. This raises several puzzles. For instance, when $q>q_\text{c}$ we expect a type I algebra with $O(1)$ entropy, yet the action is $O(1/G)$.

However, the identification of $-I$ as an entropy assumes that the Euclidean saddle \eqref{5.8a} dominates the no-boundary partition function for these geometries. This is not obviously the case. Indeed, because the action of this wormhole is negative, there are Euclidean solutions with action unbounded from below if one takes $0<H\tau<n\pi$. This leads to an unbounded path integral.
These solutions dominate even over the sphere geometry if $q$ is a dynamical variable, creating a tension with interpreting the sphere as the dominant contribution to the de Sitter entropy. It would be interesting if this could be addressed via a better understanding of the gravitational path integral. This would be analogous to the pathological vacuum solutions of a chain of spheres that were ruled out by the Kontsevich-Segal-Witten criterion \cite{Kontsevich:2021dmb,Witten:2021nzp}. 

There is also a second Lorentzian continuation of \eqref{5.8a} starting from the minimal surface $H\tau=\pi/2+\i t$. This is a cosmology with a big bang and big crunch that occurs at finite proper time, but infinite conformal time. Because of these singularities, matter correlation functions will not satisfy KMS, as explained in section \ref{sect:5.2constructive}. Indeed, one finds branch cuts in the conformal two-point function \eqref{conf2p_flrw} (see figure \ref{fig:weak_kms}). However, the correlation functions do satisfy the ``weak KMS'' property. We hope that this will suffice to define a trace. Clearly, there is much more to learn about the no-boundary state in general spacetimes.

\section*{Acknowledgments}
We thank Luca Ciambelli, Daniel Harlow, Luca Iliesiu, Victor Ivo, Jorrit Kruthoff, Juan Maldacena, Rob Myers, Vladimir Narovlansky, Pratik Rath, Gautam Satishchandran, and Jonathan Sorce for useful discussions. We are especially grateful to Edward Witten for numerous discussions that were integral to this work. AB was supported by the Marvin L. Goldberger Member Fund, the US DOE DE-SC0009988 and by the Ambrose Monell Foundation. JKF is supported by the Marvin L.~Goldberger Member Fund at the Institute for Advanced Study and the National Science Foundation under Grant No. PHY-2207584. EYU is supported by the J. Robert Oppenheimer Endowed Fund.

\appendix

\section{Cosmological constant as observer in JT gravity}\label{app:cosmo}

There is a simplification of the observer model \cite{Alexandre:2025rgx} which allows for very straightforward comparison with the Euclidean path integral. Instead of including a point-like observer, we can quantize an energy density
\begin{equation}
 q=\ell \Lambda\,.
\end{equation}
The WDW Hamiltonian becomes
\begin{equation}
 H_\text{WDW}=-p k-\Phi \ell+\Lambda \ell=0\,,\quad [\Lambda,T]=\i\,.
\end{equation}
The action for is that of a $U(1)$ 2D Yang-Mills theory coupled to JT gravity where the Hamiltonian is $\Lambda$ times the area of the surface. In the holographic limit ($\Phi=1/\varepsilon$ and $\Lambda$ finite), it is not difficult to compute the associated Euclidean wormhole path integral\footnote{We used the fact that $\Lambda A(\Sigma)$ in JT can be rewritten using the Gauss-Bonnet theorem as proportional to $\Lambda K$. The net effect is the JT finite cutoff trumpets \eqref{eq:wf_L} with $\Phi\to \Phi-\Lambda$ (as we also see from the WDW equation). However, in the holographic limit, the $\Lambda$ dependence drops out.} 
\begin{equation}
 \bra{\beta_1,\Lambda_1}\rho_\text{NB}\ket{\beta_2,\Lambda_2}=\delta(\Lambda_1-\Lambda_2)\int_0^\infty \d b\, b\,Z_\text{trumpet}(b,\beta_1) Z_\text{trumpet}(b,\beta_2)\,.
\end{equation}
It is simple to show that this is consistent with inserting the identity matrix in canonical quantization
\begin{equation}
 \mathbb{1}_\text{phys}=\int_0^\infty \d b\, b\int_{-\infty}^{+\infty}\d \Lambda \ket{b,\Lambda}\bra{b,\Lambda}.
\end{equation}
To show that this is the correct identity operator we recognize the WDW Hamiltonian as implementing the Schr\"odinger equation
\begin{equation}
 \frac{1}{\ell}\frac{\d}{\d \Phi}\frac{\d}{\d \ell}+\Phi=\i \frac{\d }{\d T}
\end{equation}
Using a separation of variables, we maydiagonalize
\begin{equation}
 \i \frac{\d}{\d T}=\Lambda.
\end{equation}
The remaining WDW equation can be solved by recognizing that we can absorb the $\Lambda$ eigenvalue upon shifting $\Phi$ by a constant. The wavefunctions are therefore
\begin{equation}
 \braket{b,\Lambda|\ell,\Phi,T}= e^{-\i T \Lambda}\frac{1}{\sqrt{2\pi}}\frac{1}{\sqrt{b^2-\ell^2}}e^{\i (\Phi-\Lambda) \sqrt{b^2-\ell^2}}\,.\label{a.7}
\end{equation}
We can compute the Klein-Gordon inner product by gauge-fixing to slices $\ell=0$ (other options are possible and give the same answer, like gauge-fixing to $T=T_0$ which gives a basis of square integrables on $\ell,\Phi$). This is the same inner product as in \eqref{3.10ip} but now with a $T$ integral. The $T$ integral produces $\delta(\Lambda_1-\Lambda_2)$. We can then shift $\Phi$ by $\Lambda$ in the $\Phi$ integral \eqref{3.10ip} to obtain
\begin{equation}
 \braket{b_1,\Lambda_1|b_2,\Lambda_2}=\frac{1}{b_1}\delta(b_1-b_2)\delta(\Lambda_1-\Lambda_2)\,.
\end{equation}
This result in the aforementioned identity operator on the physical Hilbert space, which is consistent with the Euclidean path integral. Indeed, in the asymptotic limit \eqref{a.7} is the JT trumpet \eqref{eq:trumpet}.

\bibliographystyle{ourbst.bst}
\bibliography{main}

\end{document}